\documentclass[aps,showpacs,superscriptaddress,twocolumn]{revtex4}
\usepackage[intlimits]{amsmath}
\usepackage{amssymb}
\usepackage{graphicx}
\usepackage{booktabs}
\usepackage{mathrsfs,slashed}
\usepackage{hyperref}
\usepackage{epsfig}
\usepackage{color}

\newcommand{\bea}{\begin{eqnarray}}
\newcommand{\eea}{\end{eqnarray}}

\def\gsim{\stackrel{\scriptstyle >}{\phantom{}_{\sim}}}

\usepackage{soul}

\begin{document}
	\title{Hybrid equation of state with pasta phases and third family of compact stars}
	
\author{K. Maslov}
	\email{maslov@theor.mephi.ru}
	\affiliation{National Research Nuclear University (MEPhI),
		Kashirskoe Shosse 31,
		115409 Moscow, Russia}
	\affiliation{Bogoliubov Laboratory for Theoretical Physics,
	Joint Institute for Nuclear Research,
	Joliot-Curie street 6,
	141980 Dubna, Russia}
	
\author{N. Yasutake}
	\email{nobutoshi.yasutake@p.chibakoudai.jp}
	\affiliation{Department of Physics, Chiba Institute of Technology (CIT),
		2-1-1 Shibazono, Narashino, Chiba, 275-0023, Japan}

\author{D.~Blaschke}
	\email{blaschke@ift.uni.wroc.pl}
	\affiliation{National Research Nuclear University (MEPhI),
		Kashirskoe Shosse 31,
		115409 Moscow, Russia}
	\affiliation{Bogoliubov Laboratory for Theoretical Physics,
		Joint Institute for Nuclear Research,
		Joliot-Curie street 6,
		141980 Dubna, Russia}
	\affiliation{Institute of Theoretical Physics, 
		University of Wroclaw, 
		Max Born place 9, 
		50-204 Wroclaw, Poland}

\author{A. Ayriyan}
	\email{ayriyan@jinr.ru}
	\affiliation{Laboratory for Information Technologies,
		Joint Institute for Nuclear Research,
		Joliot-Curie street 6,
		141980 Dubna, Russia}
	\affiliation{Computational Physics and IT Division, 
	A.I. Alikhanyan National Science Laboratory, 
	Alikhanyan Brothers street 2, 
	0036 Yerevan, Armenia}
	
\author{H. Grigorian}
	\email{hovikgrigorian@gmail.com}
	\affiliation{Laboratory for Information Technologies,
	Joint Institute for Nuclear Research,
	Joliot-Curie street 6,
	141980 Dubna, Russia}
	\affiliation{Computational Physics and IT Division, 
	A.I. Alikhanyan National Science Laboratory, 
	Alikhanyan Brothers street 2, 
	0036 Yerevan, Armenia}
	\affiliation{Department of Physics, 
	Yerevan State University, 
 	Alek Manukyan street 1, 
 	0025 Yerevan, Armenia}
	
\author{T.~Maruyama}
	\affiliation{Advanced Science Research Center, Japan Atomic Energy Agency, Tokai, Ibaraki 319-1195, Japan
	}

\author{T.~Tatsumi}
	\affiliation{Department of Physics, Kyoto University, Kyoto 606-8502, Japan
}
	
	\author{D. N. Voskresensky}
	\email{d.voskresensky@gsi.de}
	\affiliation{National Research Nuclear University (MEPhI),
	Kashirskoe Shosse 31,
	115409 Moscow, Russia}
	\affiliation{Bogoliubov Laboratory for Theoretical Physics,
		Joint Institute for Nuclear Research,
		Joliot-Curie street 6,
		141980 Dubna, Russia}
	
	\date{\today}
	\begin{abstract}
		The effect of pasta phases on the quark-hadron phase transition is investigated for a set of relativistic mean-field equations of state for both hadron and quark matter. The results of the full numerical solution with pasta phases are compared with those of an interpolating construction used in previous works, for which we demonstrate an adequate description of the numerical results. A one-to-one mapping of the free parameter of the construction to the physical surface tension of the quark-hadron interface is obtained for which a fit formula is given. For each pair of quark and hadron matter models the critical value of the surface tension is determined, above which the phase transition becomes close to the Maxwell construction. This result agrees well with earlier theoretical estimates.
		 The study is extended to neutron star matter in beta equilibrium with electrons and muons and is applied to investigate the effect of pasta phases on the structure of hybrid compact stars and the robustness of a possible third family 
solution.
	\end{abstract}
	\maketitle
	
	
	\section{Introduction}
	
	The recent discovery of pulsars with precisely measured masses close to $2$ M$_\odot$ such as PSR J0348+0432 \cite{Antoniadis:2013pzd} and PSR J0740+6620 \cite{Cromartie:2019kug} 
provides a new observational constraint on the equation of state (EoS) of cold dense matter under compact star (CS) conditions of $\beta-$equilibrium and global charge neutrality. 
It allowed excluding many models of CS matter for which the EoS is too soft to describe pulsars with a mass as high as $2$ M$_\odot$.
A new quality in the quest for the high-density EoS will be achieved when the NICER experiment \cite{NICER1} on board 
of the International Space Station will provide an accurate radius measurement with a 0.5 km uncertainty scale of the nearest millisecond pulsar PSR J0437-4715 with a known mass of $1.44\pm 0.07$ M$_\odot$ \cite{Reardon:2015kba}.
Further constraints on the EoS are provided by the measurement of gravitational waves from the inspired phase of the binary CS merger GW170817 \cite{TheLIGOScientific:2017qsa}.
		
	In this context it is interesting to study various possible phase transitions in the strongly interacting CS medium. 
	One of them is the transition from the hadronic to the deconfined quark matter phase. 
Its possible description and effects on the CS mass-radius diagram were studied by many authors, see \cite{Blaschke:2018mqw} and references therein. 
One of the potentially observable phenomena is the existence of a third CS branch in the CS mass-radius diagram, disconnected from that of neutron stars (NSs).
Such a third family branch can exist if there is a strong first-order phase transition \cite{Gerlach:1968zz}, e.g.,
from hadrons to quarks with a sufficiently large jump in the energy density taking place inside the CS \cite{Benic:2014jia}. 
A robust observation of pulsars with similar masses and substantially different radii (CS twin configurations) would reveal the existence of the first-order phase transition at zero temperature and thus prove the existence of the QCD critical endpoint \cite{Blaschke:2013ana}.
	
	One of the features of the first order hadron-quark phase transition is the appearance of the finite-size structures. They can appear in the CS matter due to existence of two separately conserved charges: baryon number and electric charge, 
for which the corresponding chemical potentials have to be equal in both phases to satisfy the Gibbs conditions (GC) for the phase equilibrium. 	
The electric charge has to be globally equal to zero for gravitationally self-bound objects like NSs. 
Reference~\cite{Glendenning:1992vb} suggested the presence of a wide region of mixed phase at any first-order phase transition  in multicomponent systems of charged particles; cf. further discussion of this subject  in \cite{Glendenning}.
If the global electric neutrality and the existence of the surface tension between hadron and quark matter are taken into account, then the ground state of the mixed phase consists of the finite-size droplets of one phase inside another of various geometries and sizes which has been dubbed "pasta phase".
 As has been demonstrated  in \cite{Heiselberg:1992dx},  for the appearance of the structured mixed hadron-quark phase the Coulomb plus surface energy per droplet of the new phase should have a minimum, as a function of the droplet size.  
Charge screening effects were disregarded.  
However, as was shown in \cite{Voskresensky:2001jq,Voskresensky:2002hu}, taking into account the charge rearrangement  due to the charge screening in the pasta phases has a large effect on the mixed hadron-quark phase.  
In particular, for a given pair of hadron and quark EoSs there exists such a critical value of the surface tension parameter $\sigma_c$, governed by the charge screening effects, that for any $\sigma > \sigma_c$  the resulting mixed phase will be given by the Maxwell construction  (MC)  case.  Reference \cite{Yasutake:2014oxa} argued for the importance of taking into account finite-size effects at the hadron-quark transition in  heavy hybrid stars with masses as high as $M\gsim 2 M_{\odot}$.	

The presence of structures in the mixed phase in the phase transition region results in a blurring of the energy jump, which 
may lead to the disappearance of the third family. Also it leads to an increase of the pressure $P(\mu_c)$ over the constant value $P_c$ of the MC, which can be characterized by a relative pressure excess $\Delta_P = \Delta P (\mu_c) / P_c$, where $\mu_c$ is the baryon chemical potential corresponding to the MC. 
In order to estimate the impact of the mixed phase existence on the high-mass twin (HMT) phenomenon, in 
\cite{Ayriyan:2017nby} a phenomenological approach \cite{Ayriyan:2017tvl} was used to mimic the deviation of the pressure from the MC due to the structure formation; see also Ref.~\cite{Abgaryan:2018gqp}.
It was found that the third family branch of hybrid stars joins the branch of hadronic neutron stars so that mass twin stars  cease to exist, if the pasta phase leads to $\Delta_P > 0.05$, for some models this occurs already at $\Delta_P > 0.02$ 
(see, e.g., Refs.~\cite{Ayriyan:2017nby,Blaschke:2019}). 

The purpose of the present work is to investigate the correspondence between the phenomenological construction used before and the properties of an actual pasta calculation. 
In order to do this, we compute numerically the EoS of hybrid star matter with pasta structures for a set of surface tension values $\sigma$ and then find a correspondence between $\sigma$ and $\Delta_P$. As an input we use recently developed relativistic mean-field (RMF) EoSs for hadron \cite{Maslov:2015wba} and quark matter \cite{Kaltenborn:2017hus}, labeled as KVORcut and SFM, respectively. 
Each of them contains a free parameter, allowing one to change their stiffness and thus to control the features of the mass-radius (M-R) diagram. 
We choose two hadronic and two quark parametrizations, such that all the pairs of hybrid EoSs 
pass the $2 \, M_\odot$ constraint for the maximum CS mass. 
After obtaining the $\Delta_P(\sigma)$ relation for all the combinations of the models, we compare it with the analytical expression and provide a fit formula for this key result of the present paper.

With this prerequisite we reexamine the question of the robustness of third family 
branches in the mass-radius relation of hybrid compact stars.
We discuss the model dependence and the maximum possible impact of pasta phases on other observable CS properties,
such as the moment of inertia and the tidal deformability.

	\section{Hybrid equation of state with pasta phases}
	
	In this section, we outline the description of a quark-hadron hybrid EoS with structures in the mixed phase (so-called "pasta phases") following Ref.~\cite{Yasutake:2014oxa} and  describe the recently proposed effective mixed phase construction \cite{Ayriyan:2017nby,Ayriyan:2017tvl} with the parameter $\Delta_P$ that may in turn be related to the value of the surface tension $\sigma$. 
	The input to the pasta phase calculations is the energy per particle in the hadron (H) and quark (Q) matter phases, respectively. 
	Both these functions are parameterized as
	\begin{align}
	{E}^{(H,Q)}(n_B, \beta) &= {E}_{\rm sym}^{(H,Q)} (n_B) + \beta_{(H, Q)}^2 {E}_{\rm asym}^{(H, Q)}(n_B), \,\,
	\end{align}
	where ${E}_{\rm sym}^{(H,Q)}$ and ${E}_{\rm asym}^{(H,Q)}$ are the energies per particle in symmetric matter and the asymmetry energy, respectively, in the hadron ($H$) and quark ($Q$) matter phases, and the asymmetry parameters for hadron and quark matter are defined as 
	\begin{gather*}
	\beta_{(H)}=1-2 \frac{n_p}{n_B},\\	\nonumber
	\beta_{(Q)} = \frac{n_d - n_u}{n_d + n_u} = \frac{n_d - n_u}{3 n_B}~.
	\end{gather*}
	As the microscopical input to the pasta phases code, we provide polynomial fit formulas for these EoS.
\begin{table*}[thb]
	\centering
	\caption{Parameters of the hadronic EoS fits given by Eq. (\ref{eq::fit_H}), in units of MeV.}
	\footnotesize
	\begin{tabular}{|c*{11}{|c}|}
		\hline
		&$a_0$ & $a_1$ & $a_2$ & $a_3$ &$a_4$ & $a_5$ & $a_6$ & $a_7$ & $a_8$ & $a_9$ & $a_{10} \cdot 10^3$ \\\hline
		
		KVORcut02 & 0.81351 & -34.683 & 33.997 & -56.44 & 72.585 & -47.173 & 18.49 &  -4.7982 & 0.8607  &  -0.10787 & 9.2105 \\ 
		
		KVORcut03 & 1.3516 & -36.250 & 19.305 & 7.0831 & -16.110 & 13.457 & -6.3007 & 1.8621 & -0.36858 &  0.049935 & -4.5546 \\ \hline

		& & $a_{11} \cdot 10^4$ & $a_{12} \cdot 10^6$ & $a_{13} \cdot 10^7$ & $a_{14} \cdot 10^8$ & $a_{15}\cdot 10^{10}$ & $a_{16} \cdot 10^{11}$ & $a_{17}\cdot 10^{12}$ & $a_{18}\cdot 10^{13}$ & $a_{19}\cdot 10^{15}$ & $a_{20}\cdot 10^{17}$ \\\hline
		
		KVORcut02 & & -4.8153 & 8.2274 & 7.5668 & -5.4597 & 3.6447 & 12.955 & -8.1609 & 2.4357 & -3.8209 & 2.5408 \\ 
		
		KVORcut03 & & 2.5440 & -5.0889 & -3.7728 & 3.0135 & -2.8021 & -6.9601 & 4.5957 & -1.4064 & 2.2473 & -1.5170 \\ \hline 
		
		\hline
		&$b_0$ & $b_1$ & $b_2$ & $b_3$ &$b_4$ & $b_5$ & $b_6$ & $b_7$ & $b_8$ & $b_9 \cdot 10^3$ & $b_{10} \cdot 10^4$ \\\hline
		
		KVORcut02 & 0.48339 & 47.238 & -29.873 & 21.027 & -8.8474 & 2.1619 & -0.23286 &  -0.027407 & 0.014986  &  -2.7789 &  2.9860 \\ 
		
		KVORcut03 & 0.75228 & 43.283 & -17.843 & 8.7129 & -4.8032 & 2.8450 & -1.2560 &  0.37458 & -0.076074  &  10.623 &  -10.016 \\ \hline
		
		& & $b_{11} \cdot 10^5$ & $b_{12} \cdot 10^6$ & $b_{13} \cdot 10^8$ & $b_{14} \cdot 10^9$ & $b_{15}\cdot 10^{11}$ & $b_{16} \cdot 10^{11}$ & $b_{17}\cdot 10^{12}$ & $b_{18}\cdot 10^{14}$ & $b_{19}\cdot 10^{16}$ & $b_{20}\cdot 10^{18}$ \\\hline
		
		KVORcut02 & & -1.8742 & 0.45587 & 2.5343 & -2.3493 & 2.9652 & 0.52260 & -0.36619 & 1.1538 & -1.8825 & 1.2923 \\ 
		
		KVORcut03 & & 5.8272 & -1.2946 & -8.1827 & 7.1581 & -8.4830 & -1.6147 & 1.1263 & -3.5671 & 5.8735 & -4.0798 \\ \hline 
	\end{tabular}
	\label{tab:eosH}
\end{table*}

	\subsection{Hadronic phase}
	The description of the hadronic matter phase is based on a RMF model with hadron masses and couplings dependent on the scalar field $\sigma$ developed in \cite{Kolomeitsev:2004ff, Maslov:2015wba}. Within this approach all the hadron effective masses decrease in the medium with the same rate as functions of the mean $\sigma$ field, in accordance with the idea of  the partial chiral symmetry restoration.
Phenomenological scaling functions enter the EoS only in combinations 
	\begin{gather}
		\eta_M(\sigma) = \frac{\Phi_M^2(\sigma)}{\chi_M^2(\sigma)},
	\end{gather}
	where the subscript $M = \sigma, \omega, \rho, \phi$ labels the meson fields included into the model. In this framework the KVORcut family of the models was constructed, which allows for a high maximum CS mass and simultaneously fulfills a majority of other constraints. We focus here on the KVORcut02 and KVORcut03 models, in which the additional stiffness is introduced as outlined in \cite{Maslov:cut} to allow for the description of pulsars with a mass of $\approx 2 $ M$_\odot$ \cite{Antoniadis:2013pzd} even when hyperons and $\Delta$ resonances are present in the EoS. The KVORcut02 model is the stiffest one, while the softer KVORcut03 model passes constraints for the pressure as a function of the baryon density following from analyses of flows in heavy-ion collisions \cite{Danielewicz:2002pu, Lynch:2009vc}.  
	
	The hadronic EoS can be parametrized with
	\begin{eqnarray}
	{E}_{\rm sym}^{(H)} (n_B) &=& \sum_{i=0}^{20} a_i u^i,~
	{E}_{\rm asym}^{(H)} (n_B) = \sum_{i=0}^{20} b_i u^i, 
	\label{eq::fit_H}
	\end{eqnarray}
	where  $u={n_B}/{n_0}$ is the nuclear compression with $n_0 = 0.16 \, {\rm fm}^{-3}$ being the nuclear saturation density.
	The coefficients $a_i, \, b_i$ are the fit parameters given in Table~\ref{tab:eosH}.
	
\begin{table*}
	\caption{Parameters of the quark SFM EoS fits given by Eq.~(\ref{eq::fit_Q}).
		All units are MeV except for $\gamma_{a,s}$ which are dimensionless.}
	\begin{tabular}{|c|c|c|c|c|c|c|c|}
		\hline
		& $\alpha_s$ & $\beta_s$ & $\gamma_s$ & $\delta$ & $\alpha_a$ & $\beta_a$ & $\gamma_a$ \\
		\hline
		$\alpha=0.2$& -1.7641 & 1.6654 & 0.33419 & 0.98796 & 
		-0.02570 & 0.052563 & 0.0085143 \\
		$\alpha=0.3$ &	-6.8814 & 3.3411 & 4.5326 & 0.21680	& -0.029531 & 0.056170 &  0.018573	\\	\hline
	\end{tabular}
	\label{tab:eosQ}
\end{table*}

	\subsection{Quark matter phase}
	For description of the quark phase we use the recently developed RMF density functional \cite{Kaltenborn:2017hus}, inspired by the string-flip model \cite{Horowitz:1985tx, Ropke:1986qs}. This approach gives a simple way to model the confinement of quarks via introducing divergent quark masses for low baryon densities and the density-dependent screening effect. The effective screening is described analogously to the excluded volume effect in models of hadronic matter with density-dependent couplings.
	
	Together with these features, the model incorporates a repulsive vector interaction with a higher-order density dependence. It gives a relatively soft EoS near the phase transition point, which becomes much stiffer as the density increases. This stiffness allows for description of the $2 M_\odot$ CS mass constraint for hybrid stars. 
	However, the soft behavior near the phase transition (PT) leads to the possibility of mass twin CS configurations, which 
appear if the density jump is large enough for the existence of a separate branch of hybrid stars dubbed the "third family" of CSs. Further details on these models can be found in \cite{Kaltenborn:2017hus}.                     
	
	This EoS can be parametrized using 
	\begin{eqnarray}
	{E}_{\rm sym}^{(Q)}(n_B) &=& \alpha_s + \beta_s  u^{-1/3} + \gamma_s  u^{1/3} + \delta \, u, \nonumber \\ 
	{E}_{\rm asym}^{(Q)}(n_B) &=& \frac{\alpha_a u} {1 + \gamma_a u^2} + \beta_a u^{2/3},
	\label{eq::fit_Q}
	\end{eqnarray}
	where $\alpha_{a,s}, \beta_{a, s}, \gamma_{a,s}$ and $\delta$ are the fit parameters
	given in Table~\ref{tab:eosQ}.

	\subsection{Pasta phase calculation}
	
	In order to study the finite-size structures, we require the GC 
	\begin{gather}
		P^{(H)} = P^{(Q)}, \quad \mu_B^{(H)} = \mu_B^{(Q)}, \quad \mu_e^{(H)} = \mu_e^{(Q)}
	\end{gather}
	to be fulfilled within the mixed phase region. In the Wigner-Seitz (WS) approximation the space is considered to be tesselated by cells of a given geometry depending on the dimensionality $d$: spheres for $d = 3$, cylinders for $d = 2$, and slabs for $d = 1$ with a volume $V_W$. Within each of the cells the quark phase of the volume $V_Q$ is embedded into the surrounding hadron phase of the volume $V_H$ or vice versa. If the hadron phase is dominant the structures are dubbed "droplets" and "rods" for $d = 3$ and $d = 2$, respectively, and "bubbles" or "tubes" otherwise. 
	The boundary layer is assumed to be smaller than any of the characteristic lengths of the problem, and thus the surface effects are captured by introducing the surface tension parameter $\sigma$. The value of $\sigma$ is highly model dependent and uncertain. In this work we vary $\sigma$ in a wide range with a small step, which allows us to determine numerically the critical surface tension $\sigma_c$ for all pairs of hadron and quark models under consideration.
	
	In the Thomas-Fermi approximation the Helmholtz free energy of a cell reads
	
	\begin{align}
	\epsilon &= \int_{V_H} d^3 r {\cal E}^{(H)}[\{n_h(r)\}] + \int_{V_Q} d^3 r {\cal E}^{(Q)}[\{n_q(r)\}] + \epsilon_e \nonumber \\ &+ \epsilon_C + \epsilon_S,
	\end{align}
	where $h = n,p$ and $q = u,d$; ${\cal E}^{(H)} = n_B(E^{(H)} + m_N)$ and ${\cal E}^{(Q)} = n_B E^{(Q)}$ are the free energy densities of hadron and quark matter, respectively, and  $\epsilon_e$, $\epsilon_C$, $\epsilon_S$ stand for the contributions to the energy per cell from free electron gas, Coulomb effects, and the surface term, respectively. The nucleon mass is $m_N = 938$ MeV. 
First in Sec.~\ref{sec:Numerics} we will consider the formation of pasta without taking muons into account, and then in 
Sec.~\ref{sec:muons} we will study the impact of their contribution on the pasta properties.

The Coulomb contribution to the energy per cell is given by
	\begin{align}
		\epsilon_C = \frac{e^2}{2} \int_{V_{WS}} d^3 r d^3 r' \frac{n_{\rm ch}(\vec r) n_{\rm ch}(\vec r \, ')}{|\vec r-\vec r\,'|},
	\end{align}
	where the charge density is
	\begin{align}
		e n_{\rm ch}(\vec r) = \sum_{h} Q_h n_h(\vec r) + \sum_{q} Q_q n_q(\vec r) - e (n_e+ n_\mu).
	\end{align}
	Accordingly, the screened Coulomb potential $\phi(r)$ is defined as 
	\begin{equation}
	\phi(r)=-\int_{V_{WS}} d^3 r' \frac{e^2\, n_{\rm ch}(\vec r \, ')}{|\vec r-\vec r\,'|} +\phi_0~,
	\end{equation}
	where $\phi_0$ is an arbitrary constant representing the gauge degree of freedom. 
	It is fixed by the condition $\phi(R_{WS})=0$; see Ref.~\cite{Yasutake:2014oxa}.
	The self-consistent field fulfills the Poisson equation
	\begin{equation}
	\Delta \phi(r) = 4\pi e^2 n_{\rm ch}(r)~.
	\end{equation}
	The equations are solved for a given baryon density
	\begin{align}
		n_B = \frac{1}{V_{WS}} \left[\sum_{h} \int_{V_H} d^3 r \, n_h(r) 
		 + \frac{1}{3} \sum_{q} \int_{V_Q} d^3 r \, n_q (r)\right]
	\end{align}
	together with the charge neutrality condition
	\begin{align}
	\int_{V_{WS}} d^3 r \, n_{\rm ch} (r) = 0. 
	\end{align}

	\begin{figure}
	\includegraphics[width=\linewidth]{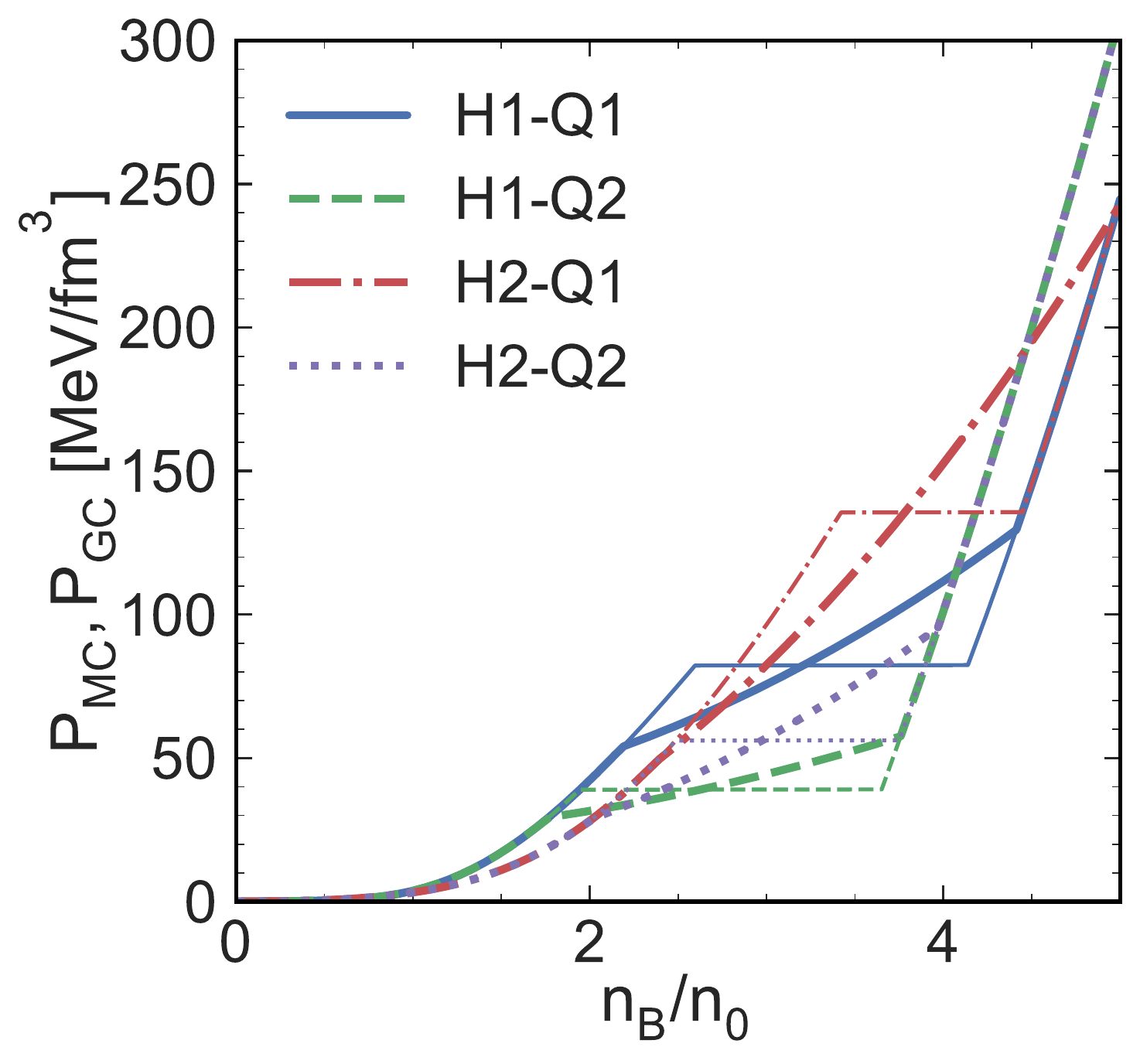}
	\caption{Pressure as a function of the density for all the combinations of models under consideration with the MC (thin lines) and the GC with $\sigma=0$ and no electrostatic contribution (thick lines) for the case without muons.}
	\label{fig::P_maxw_GC}
    \end{figure}

\begin{figure*} 
	\includegraphics[width=.9\linewidth]{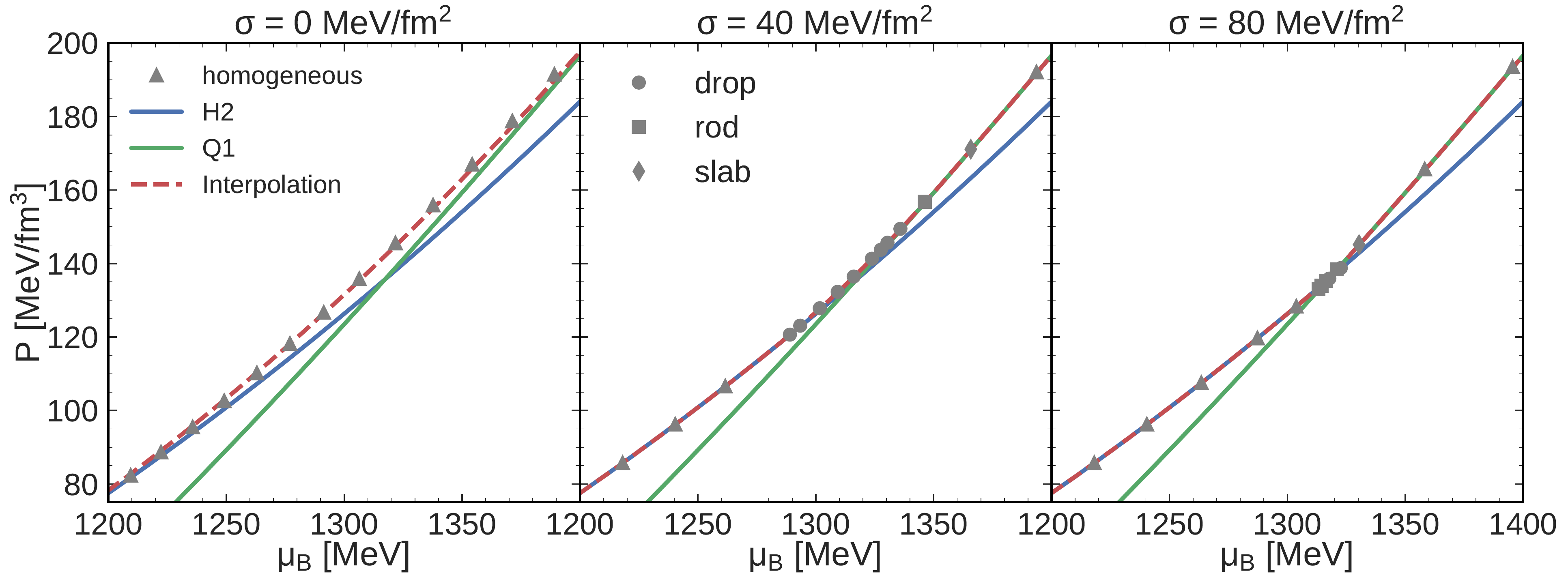}
	\includegraphics[width=0.9\linewidth]{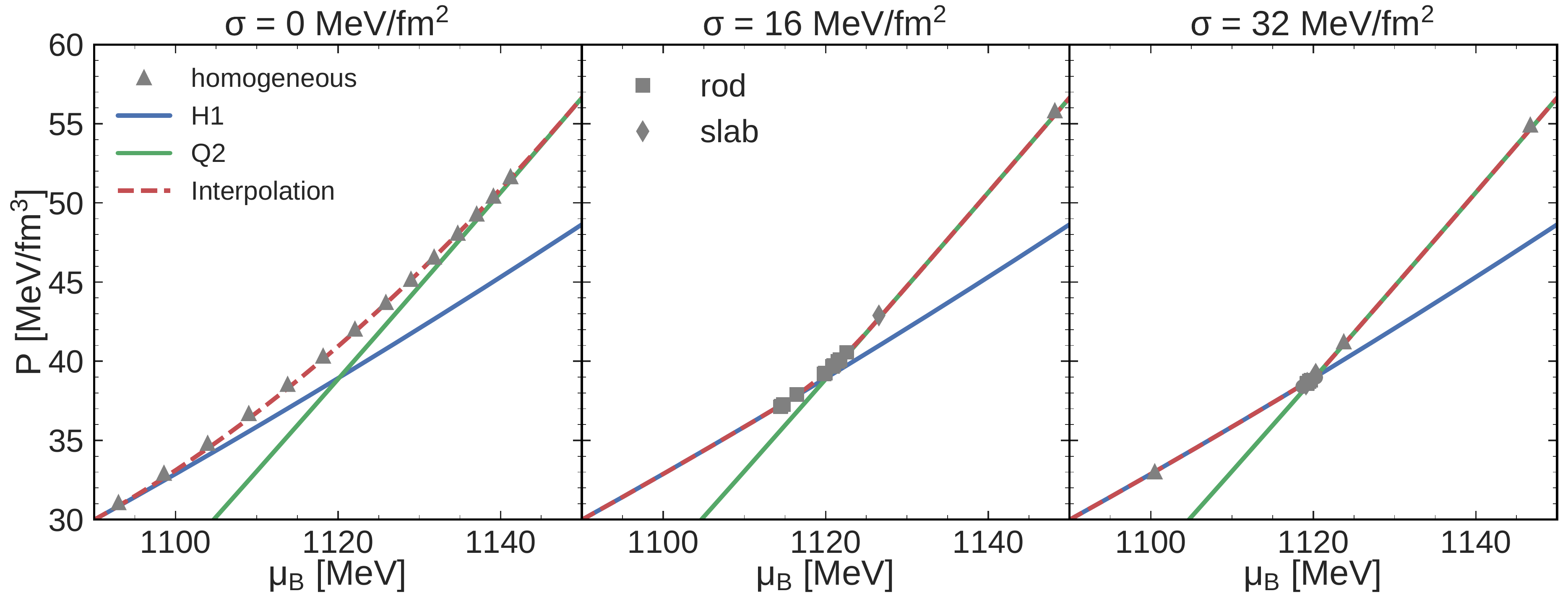}
	\caption{Pressure as a function of the chemical potential for the H2-Q1 model for the surface tension $\sigma = [0, 40, 80]$ MeV/fm$^2$
	(upper panel) and the H1-Q2 model with $\sigma = [0, 16, 32]$ MeV/fm$^2$ (lower panel). Solid lines denote the quark and hadron EoSs, 
	symbols denote the numerical results with structure types indicated in the legend, and the dashed lines show the best-fit curves using Eq.~(\ref{eq::P_grigoryan}).}
	\label{fig::fits-mu}
\end{figure*}    

\begin{figure*} 
	\includegraphics[width=0.9\linewidth]{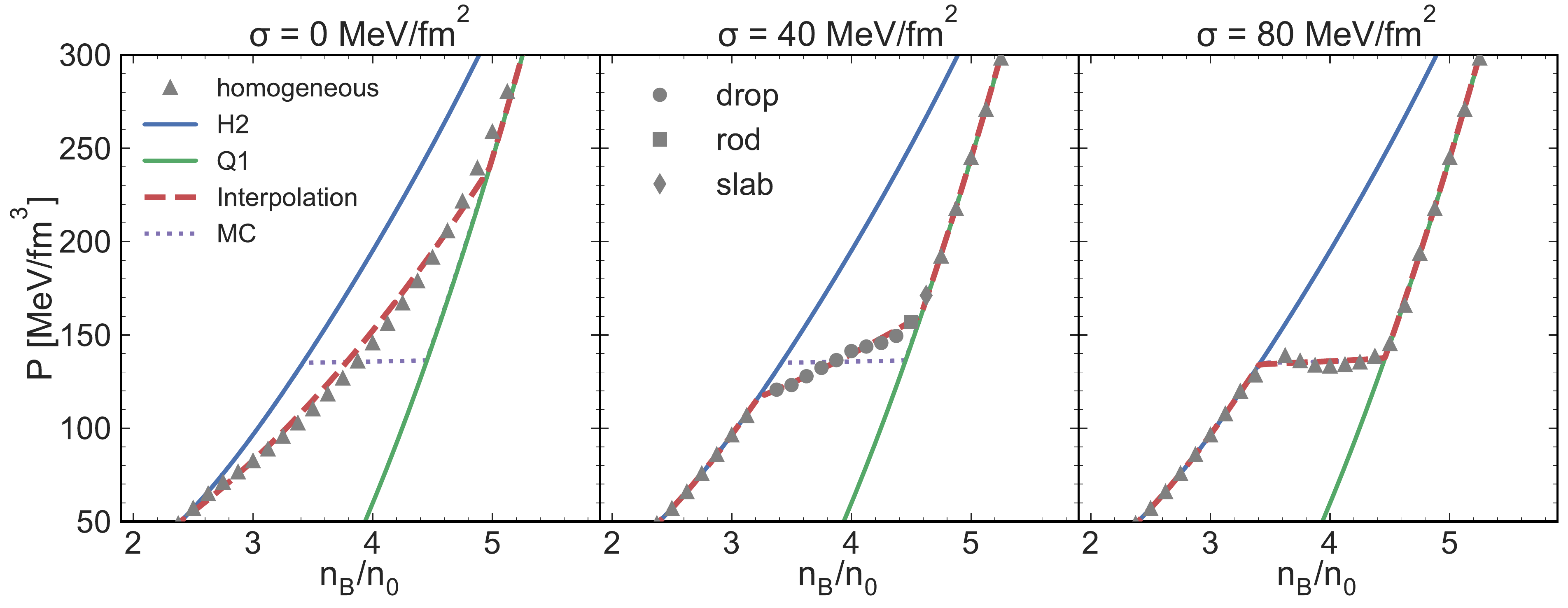}
	\includegraphics[width=0.9\linewidth]{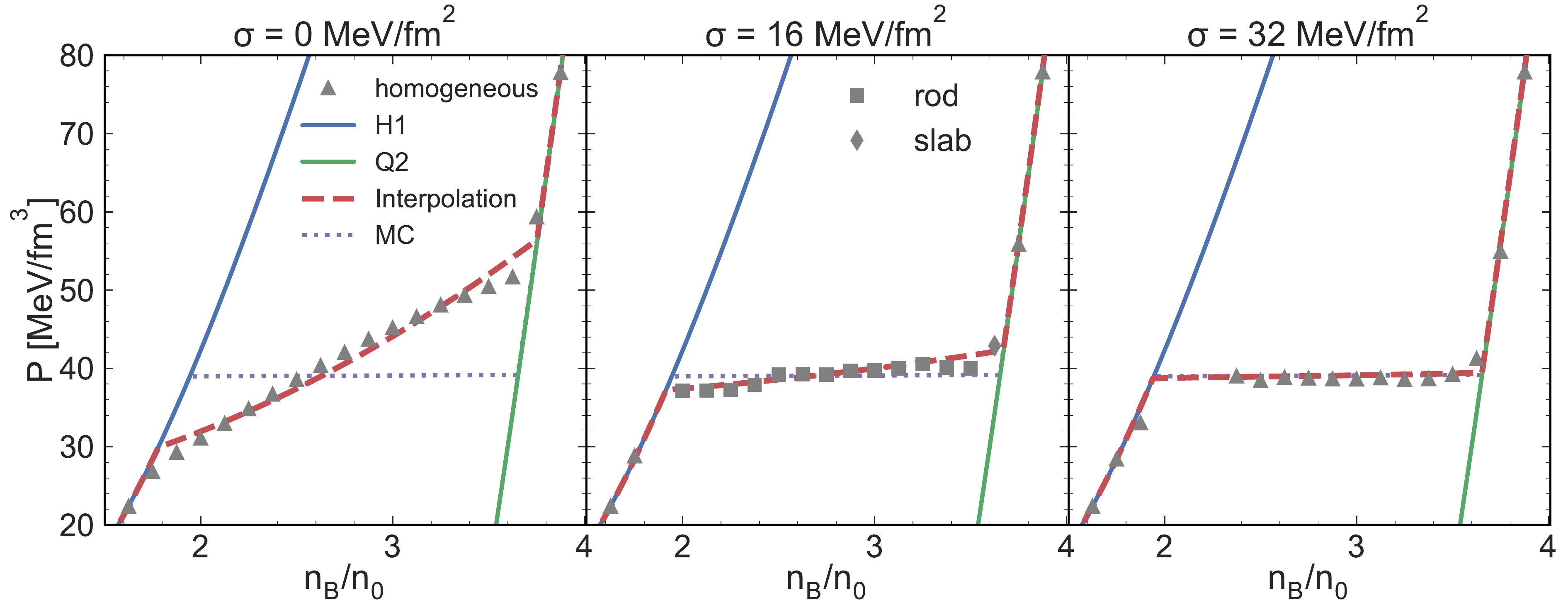}
	\caption{Pressure as a function of the baryon density for the H2-Q1 model for the surface tension $\sigma = [0, 40, 80]$ MeV/fm$^2$
	(upper panel) and the H1-Q2 model with $\sigma = [0, 16, 32]$ MeV/fm$^2$ (lower panel). 
	Solid lines denote the quark and hadron homogeneous matter EoSs, symbols denote the numerical results with the structure type indicated in the legend, and the dashed lines show the best-fit curves using Eq.~(\ref{eq::P_grigoryan}).
	For comparison, we show also the MC by a dotted line.
In the rightmost panels where $\sigma > \sigma_c$ we use the symbols for homogeneous phase also in the coexistence region because the space is divided just into two phases and no structures can be identified. 
	}
	\label{fig::fits}
\end{figure*}    

\subsection{Fit formula for the phase transition}
In \cite{Ayriyan:2017nby,Ayriyan:2017tvl} a simple modification of the MC was employed to mimic the effect of the pasta structures on the quark-hadron phase transition. If the EoSs of quark and hadron matter are given in terms of the pressure as a function of the baryon chemical potential $P^{(H)}(\mu)$ and $P^{(Q)}(\mu)$, respectively, the pressure with this construction is given by

\begin{gather}
P(\mu) = 
\begin{cases}
P^{(H)}(\mu), & \mu < \mu_{cH}, \\
P^{(M)}(\mu), & \mu_{cH} < \mu <\mu_{cQ},\\
P^{(Q)}(\mu), & \mu_{cQ} < \mu,
\end{cases}
\label{eq::P_grigoryan}
\end{gather}
where 
\begin{equation}
P^{(M)}(\mu) = a (\mu - \mu_c)^2 + b (\mu - \mu_c) + (1 + \Delta_P)P_c
\label{eq::P_mixed}
\end{equation}
is a parabolic ansatz  \cite{Ayriyan:2017nby,Ayriyan:2017tvl} for the pressure in the mixed phase 
and $P_c = P^{(H)}(\mu_c) = P^{(Q)}(\mu_c)$ is the intersection point of hadron and quark matter EoS at $\mu=\mu_c$ 
corresponding to the MC. From this expression the baryon density can be calculated as $n_B(\mu) = dP/d\mu$.
The four parameters $a$, $b$, $\mu_{cH}$, $\mu_{cB}$ can be determined from the conditions of continuity of pressure and baryon number density $n_B(\mu)$ at both $\mu = \mu_{cH}$ and $\mu = \mu_{cQ}$, so there is only one free parameter $\Delta P$ left.
For other ans\"atze interpolating between the hadronic and quark matter phases with fixed end-points, see Ref.~\cite{Abgaryan:2018gqp}.

\section{Numerical results}
\label{sec:Numerics}	
	In this section we show the results of the fits for every combination of the models. Below we use the ``\textit{H}-\textit{Q}" notation for the pairs of hadronic ($H$) and quark ($Q$) models, where $H = $ H1, H2 corresponds to KVORcut[02, 03] models, respectively, and $Q =$ Q1, Q2 denotes the SFM model with $\alpha=0.2, 0.3$, respectively.
	
For each pair of the models Fig.~\ref{fig::P_maxw_GC} demonstrates two limiting cases of the MC and the GC 
with $\sigma=0$ and no Coulomb energy 
contribution. 
For a given quark EoS the transition density and pressure are lower for a stiffer hadronic EoS H1, and for a given hadronic EoS the transition happens earlier for the softest quark EoS Q2. Thus the onset of the phase transition proves to be the lowest for the H1-Q2 case, because in this case the hadronic EoS is the stiffest one and the quark EoS is the softest in the low density region. 	
	After taking into account the continuity of the electron chemical potential, the pressure becomes non-constant within the mixed phase region. It can be seen from Fig.~\ref{fig::P_maxw_GC} that the pressure difference on the GC increases with an increase of the $P_c$ of the MC. This broadening is to be compared with the critical PT broadening from \cite{Ayriyan:2017nby} in order to understand whether the third branch phenomenon can be in principle eliminated by the pasta phases formation.

\begin{table*}[thb]
\caption{Critical pressure $P_c$, energy density ${\cal E}_c$ and energy density jump $\Delta {\cal E}$ in units of MeV/fm$^3$, together with the best fit parameters of Eq.~\ref{eq::S(x)} and the theoretical estimates for all the combinations of models without muons (upper table) and with muons (lower table). Numerical values of the critical surface tension
$\sigma_c$ and the analytic results $ \widetilde \sigma_c$ given by Eq.~(\ref{eq::sigma_c_an}) in units of MeV/fm$^2$. 
The Debye lengths $\lambda_D^{(Q)}$ and $\lambda_D^{(H)}$ are given in units of fm.}
\centering
\begin{tabular}{| c | ccc | ccc | ccc |}
\hline & \multicolumn{9}{c|}{without muons} \\
\hline
Pair & $\Delta_P(0)$ &$\sigma_c$& $\beta$ & $P_c$ & $\Delta {\cal E}$ & ${\cal E}_c$
& $\widetilde \sigma_c$ & $\lambda_D^{(Q)}$ & $\lambda_D^{(H)}$
\\ \hline
H1-Q1 & $0.053 \pm 0.001$ & $52.3 \pm 1.27$ & $0.64 \pm 0.03$ & 82 & 306 & 433  & 49.2   & 3.79 & 5.87 \\ \hline
H1-Q2 & $0.053 \pm 0.002$ & $30.2 \pm 1.90$ & $0.62 \pm 0.08$ & 39 & 306 & 309  & 25.4   & 3.96 & 6.40 \\ \hline
H2-Q1 & $0.048 \pm 0.001$ & $74.1 \pm 0.94$ & $0.77 \pm 0.05$ & 135 & 215 & 585 & 73.3  & 3.71 & 5.44 \\ \hline
H2-Q2 & $0.060 \pm 0.001$ & $38.6 \pm 1.06$ & $0.67 \pm 0.04$ & 56 & 232 & 401   & 38.5   & 3.93 & 6.00 \\ \hline
\hline & \multicolumn{9}{c|}{with muons} \\
\hline
Pair & $\Delta_P(0)$ &$\sigma_c$& $\beta$ & $P_c$ & $\Delta {\cal E}$ & ${\cal E}_c$
& $\widetilde \sigma_c$ & $\lambda_D^{(Q)}$ & $\lambda_D^{(H)}$
\\ \hline
H1-Q1 & $0.046\pm 0.001$ & $43.3 \pm 0.89$ & $0.65 \pm 0.03$ & 85 & 297 & 444  & 42.1   & 3.77 & 5.22 \\ \hline
H1-Q2 & $0.049\pm 0.002$ & $29.0 \pm 1.53$ & $0.43 \pm 0.04$ & 40 & 302 & 314  & 22.5   & 3.96 & 5.79 \\ \hline
H2-Q1 & $0.039 \pm 0.001$& $59.4 \pm 0.97$ & $0.84 \pm 0.07$ & 144 & 203 & 607 & 62.3   & 3.69 & 4.78 \\ \hline
H2-Q2 & $0.052 \pm 0.001$& $35.7 \pm 0.92$ & $0.63 \pm 0.03$ & 58 & 226 & 409 & 33.8 & 3.93 & 5.38 \\ \hline
\end{tabular}
\label{tab:Psigma_fit}
\end{table*}

	\subsection{Comparison with phenomenological description}
		
	Possible applications of the construction (\ref{eq::P_grigoryan}) require knowledge of the limits for realistic values of the parameters. The parameter $\Delta_P$ is greater than zero by definition and limited from above, with the maximum possible $\Delta P$ corresponding to zero surface tension. The exact correspondence of $\Delta_P$ to the surface tension $\sigma$ is presented in this section.

	We performed the least-squares fit of the $P(n_B)$ derived from Eq.~(\ref{eq::P_mixed}) to describe the numerical data, using $\Delta_P$ as a variational parameter. 
	We demonstrate the results of the fitting in Figs.~\ref{fig::fits-mu} and \ref{fig::fits} in terms of $P(\mu)$ and $P(n)$, respectively, for the  examples of the combinations H2-Q1 (top panels) and H1-Q2 (bottom panels). The choice of surface tension values shown there, $\sigma = [0, 40, 80]$ MeV/fm$^2$ for H2-Q1 and $\sigma = [0, 16, 32]$ MeV/fm$^2$, roughly corresponds to 
$[0, \sigma_c/2, \sigma_c]$ for each pair of models. 	
Symbols show the numerical results for the pasta structures, solid lines denote the initial hadron and quark EoSs and the best fit by formula Eq.~(\ref{eq::P_grigoryan}) is shown by the dashed line. 
We see that the phenomenological construction we employed in previous works can adequately describe the exact numerical result. 
The quality of the fit can be characterized by the root-mean-square deviation, defined as 
	\begin{align}
		\chi = \sqrt{\frac{1}{N} \sum_{i = 1}^N \Big(\frac{P_{\rm fit}(n_B^{(i)}) - P_i}{P_i}}\Big)^2,
	\end{align}
where $P_i$ and $n_B^{(i)}$ are the calculated numeric data points, $P_{\rm fit}(n_B)$ is given by Eq.~(\ref{eq::P_grigoryan}), and $N$ is the number of data points in the range of densities where the pasta structures exist. 
For the cases shown in the upper panel of Fig.~\ref{fig::fits}, the rms deviation $\chi$ equals to $3.5 \%, 1.0 \%, 1.6 \%$ for $\sigma = [0, 40, 80]$ MeV/fm$^2$, respectively. 
With such a precision the use of the interpolating construction, given by Eq.~(\ref{eq::P_grigoryan}), for mimicking the pasta phase effects on the EoS is justified. 
	
Similar fits were performed for a set of surface tension values in the range $(0 - 80) \, $MeV/fm$^2$. 
In Fig.~\ref{fig::P_sigma} we show by solid lines the so-obtained curves $\Delta_P(\sigma)$ for all the combinations of the models. It is clearly seen that if the surface tension exceeds some critical value $\sigma_c$, the effect of the finite-size structures becomes negligible. 
	However, the pressure excess $\Delta_P$ does not become exactly zero, since in the MC case the WS cell size would become infinite, but the numeric code has a large but finite limiting cell size. Thus the code we used is applicable only for $\sigma<\sigma_c$. Nevertheless, the resulting pressure in the $\sigma>\sigma_c$ is very close to the MC line, and is best fitted by a very low $\Delta_P$.

	\subsection{Model dependence of the critical surface tension}
	
\begin{figure}[!htb]
		\includegraphics[width=\linewidth]{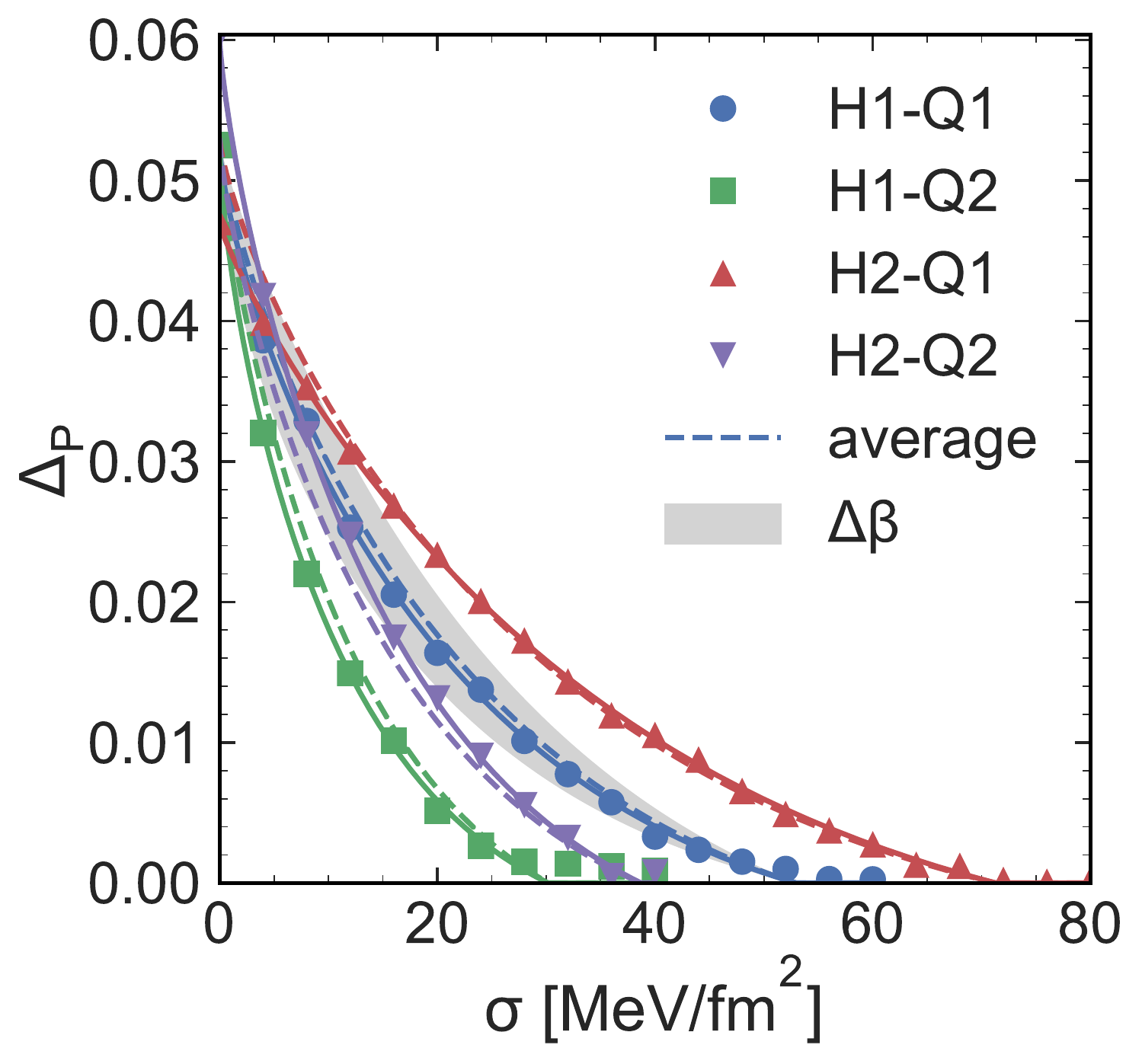}
		\includegraphics[width=\linewidth]{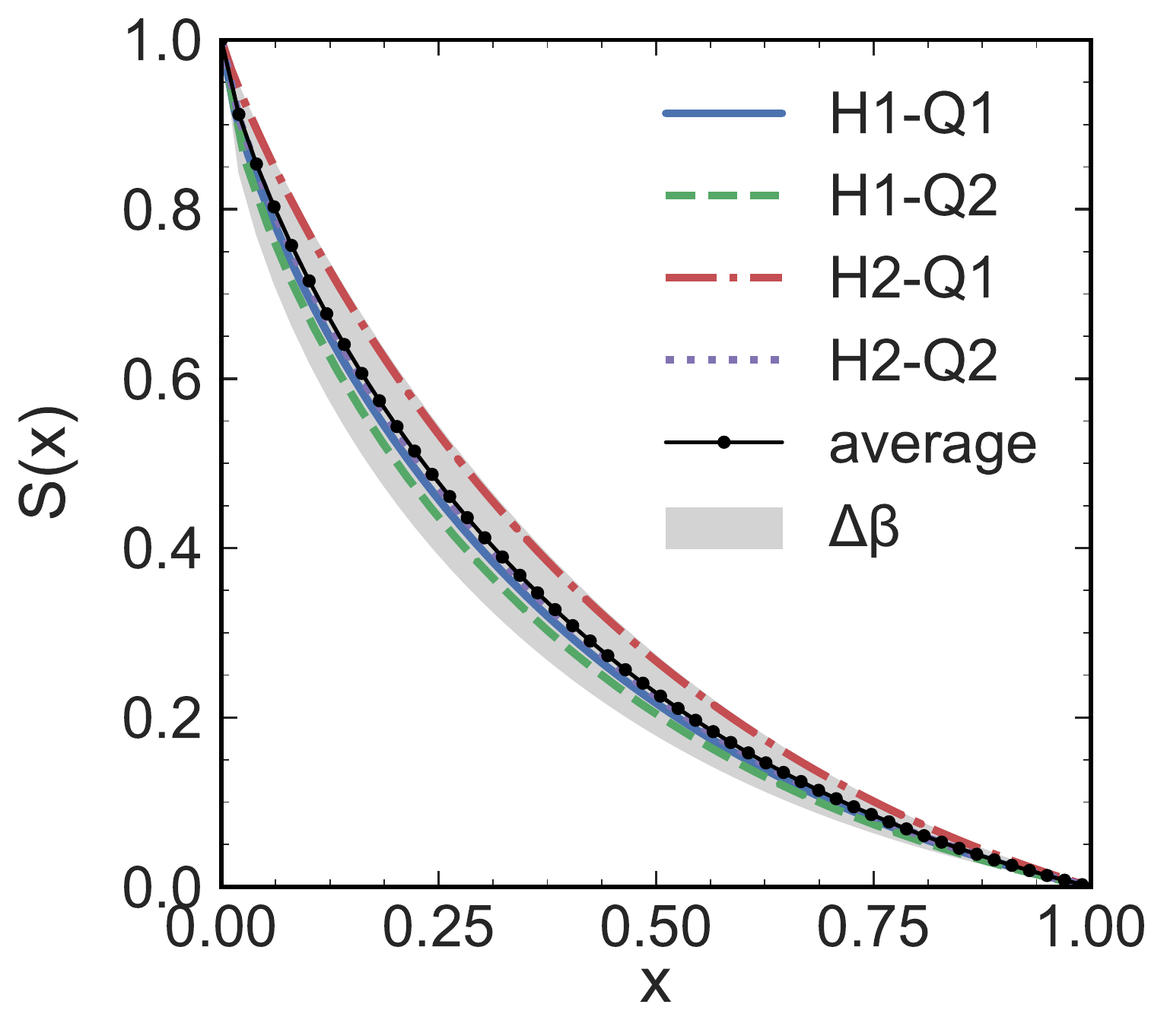}
		\caption{Upper panel: Relative pressure shift $\Delta_P$ at the MC chemical potential $\mu_c$ as a function of the surface tension 
		$\sigma$ for all the combinations of models. 
		Solid lines show the results of the fit with use of Eq.~(\ref{eq::S(x)}), and the dashed lines are plotted with use of the mean value of 
		$\beta$ (\ref{eq:mean}) and the best-fit $\Delta_P^{(0)}$ and $\sigma_c$ for each model.
		Lower panel: The dimensionless function $S(x)$ (\ref{eq::S(x)}) for our models. The solid line with dots shows  $S(x)$ for the mean 	
		value $\beta$, and the shaded area indicates the change of $S(x)$ for the variation of $\beta$ with $\Delta\beta = 0.18$.
		}
		\label{fig::P_sigma}
	\end{figure}

	Because of nonexact reproduction of the MC line in the case of large surface tension $\sigma$, we need to perform an additional step to define the critical surface tension $\sigma_c$ in a reproducible manner. For this we used a fit of the $\Delta_P(\sigma)$ curve with the following ansatz:
	\begin{eqnarray}
	\label{eq::DeltaP}
	\Delta_P(\sigma) &=& \Delta_P(0) S({\sigma}/{\sigma_c}; \beta),\\ 
	S(x; \beta) &=& e^{-x}(1 - x^\beta) \theta(1-x),
	\label{eq::S(x)}
	\end{eqnarray}
where $\Delta_P(0)$, $\beta$, and $\sigma_c$ are the parameters of the fit, and $\theta(x)$ is the step function. 
The parameter values thus determined are summarized in Table \ref{tab:Psigma_fit}. 
The solid lines in the upper panel of Fig.~\ref{fig::P_sigma} show the best fit curves given by Eq.~(\ref{eq::S(x)}). 

	The dimensionless function $S(x)$ shown in the lower panel of Fig.~\ref{fig::P_sigma} is only weakly model dependent.
By the solid line with dots we show the function $S(x; \overline{\beta})$, where 
\begin{eqnarray}
\overline{\beta} = 0.68
\label{eq:mean}
\end{eqnarray}	
is the mean value of the $\beta$ parameter. 
The region covering all the $S(x)$ for our models is given by symmetrically varying $\beta$ with ${\Delta \beta} = 0.18$. It is shown by the shaded area and labeled by $\Delta \beta$ in the legend. 
The same shaded area is plotted as a function of the dimensionful $\sigma$ in the upper panel for the H1-Q1 model as an example, with the mean value $\overline{\Delta}_P(0)=0.053$.
It is interesting to note that this maximal value for the parameter of the mixed phase construction (\ref{eq::P_grigoryan}) 
is in the same range of $5~\%$ that was found in Ref.~\cite{Ayriyan:2017nby} as the critical value above which the third family 
solution for hybrid compact stars would cease if it existed for the MC.
For the hybrid EoS example considered here this would concern the cases with H1, the stiffer hadronic EoS.  
This finding means that even for the GC with $\sigma=0$ a third family of compact stars can be obtained, as was the case for the work of Glendenning and Kettner \cite{Glendenning:1998ag} where the notion of mass twin stars was introduced. 
Our results for the properties of compact star sequences will be discussed in detail in Sec.~\ref{sec:CS} below.
	\begin{figure}[!htb]
		\includegraphics[width=\linewidth]{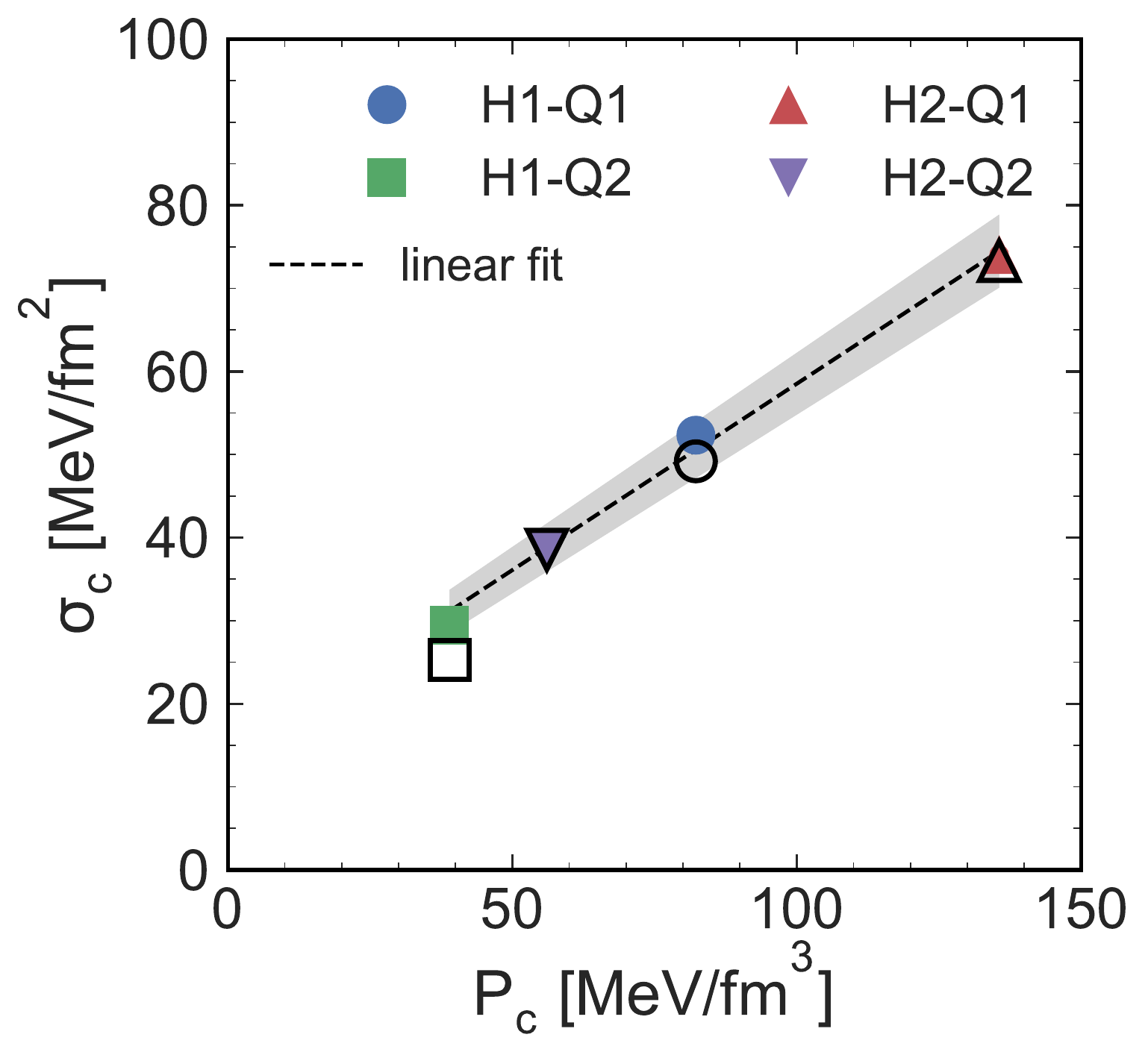}
		\caption{Critical surface tension $\sigma_c$ as a function of the pressure on the MC line $P_c$ for all the combination of models considered. Full {symbols} denote the results of the numerical calculation, and open {symbols} are the analytical estimates using Eq. \eqref{eq::sigma_c_an}.
		The dashed line shows a linear fit given by Eq.~(\ref{eq::linfit}) and the gray area denotes its estimated uncertainty.}
		\label{fig::sigma_c_Pc}
	\end{figure}
	
	In order to describe the model dependence of the critical surface tension $\sigma_c$ we choose the pressure on the MC line $P_c$ as the parameter characterizing a pair of models. We show the dependence of $\sigma_c$ on $P_c$ in Fig.~\ref{fig::sigma_c_Pc} by solid symbols. It is clear that within the current set of models the critical surface tension monotonically grows with an increase of the pressure $P_c$.	
	It is interesting to compare this dependence with the expression for $\sigma_c$ obtained in \cite{Voskresensky:2002hu} obtained analytically using the linearized Poisson equation. The resulting expression involves the Debye screening lengths and the values of the electric field in both phases. At the critical chemical potential $\mu_B = \mu_{c, B}^{(H)}$ of the phase transition we have $\mu_{e, \rm Gibbs} = \mu^{(H)}_{e,\rm bulk}$. When the quark matter fraction is small, the size of the hadron phase is  much larger than the screening length. Thus we can consider it to be electrically neutral with $\mu_e = \mu_{e, \rm bulk}^{(H)}$. The electron contribution to the quark matter charge can be neglected \cite{Voskresensky:2002hu}. Hence all the values are to be evaluated at $\mu_B = \mu_{c, B}^{(H)}$ and $\mu_e = \mu_{e, \rm bulk}^{(H)}$ for hadronic matter and $\mu_e = 0$ for quark matter. The Debye screening lengths are 
	\begin{equation}
	\left[\lambda_D^{(H, Q)}\right]^{-2} = - 4 \pi e^2 \frac{\partial n_{\rm ch}^{(H,Q)}}{\partial \mu_e}\Bigg|_{\mu_B},
	\label{eq::lambda_D}
	\end{equation}
	where $e^2 = 1/137$ and $n_{\rm ch}^{(H,Q)}(\mu_B, \mu_e)$ is the charge density of the phase $H$ or $Q$. 
They can be expressed through Landau parameters of beta-equilibrium matter, though in this work we calculated the derivatives numerically. 
The expression for the critical surface tension is (see Eq.~(77) of Ref.~\cite{Voskresensky:2002hu})
%
%
\begin{equation}
\widetilde{\sigma}_c =  \frac{(U_0^{\rm II}-U_0^{\rm I})^2}{8\pi e^2 (\lambda_D^{(H)} + \lambda_D^{(Q)})}~,
\label{eq::sigma_c_an}
\end{equation}
where the parameters of the electric field contribution are $U_0^{\rm II} \simeq -\mu_{e, \rm bulk}^{(H)}$ and 
$U_0^{\rm I} = - 4 \pi e^2 (\lambda_D^{(Q)})^2 n_{\rm ch}^{(Q)} (\mu_B=\mu_c, \mu_e = 0)$.
	
	The so-obtained values of the parameters and $\widetilde{\sigma}_c$ are summarized in Table \ref{tab:Psigma_fit}. 
In Fig.~\ref{fig::sigma_c_Pc} we show by the open symbols the theoretical values of the critical surface tension $\widetilde \sigma_c$ from Eq.~(\ref{eq::sigma_c_an}) against the pressure $P_c$  of the MC. 
We see that these values are consistent with those obtained from fitting the numerical result. 
Thus we confirm that the numerical calculation correctly captures the essential physics of the pasta phase formation and proves the usability of the linearized Poisson equation for the description of the electric field.

We also show by the dashed line the linear fit to these data points,
\begin{equation}
\sigma_c = d \, (P_c - P_0) + \sigma_0~,
\label{eq::linfit}
\end{equation}
where $\sigma_0=30.3 \pm 0.83$ MeV/fm$^2$, , $P_0=40$ MeV/fm$^3$, and $d=0.28 \pm 0.01$ fm.
Equation~(\ref{eq::linfit}) is a key result of this work. 
This relationship, together with Eqs. (\ref{eq::DeltaP}), (\ref{eq::S(x)}), and the mixed phase construction
(\ref{eq::P_grigoryan}), (\ref{eq::P_mixed}) allows one to obtain the family of hybrid equations of state equivalent to 
those of the pasta phase construction, with the surface tension $\sigma$ as a free parameter, just from the knowledge
of two EoSs for the pure phases which define the critical pressure $P_c$ of their MC. 

\begin{figure}[!htb]
	\centering
	\includegraphics[width=0.95\linewidth]{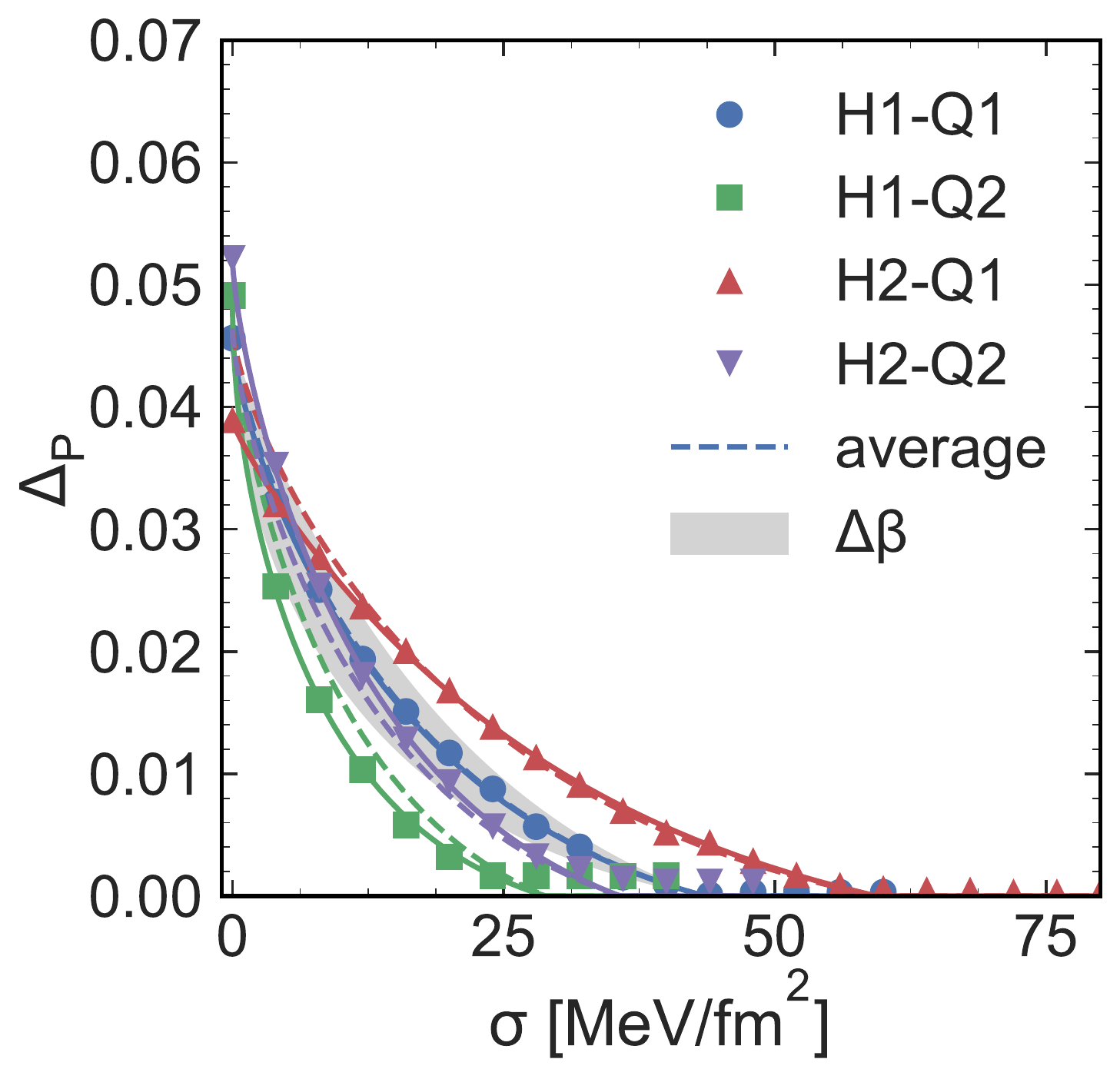}
	\includegraphics[width=\linewidth]{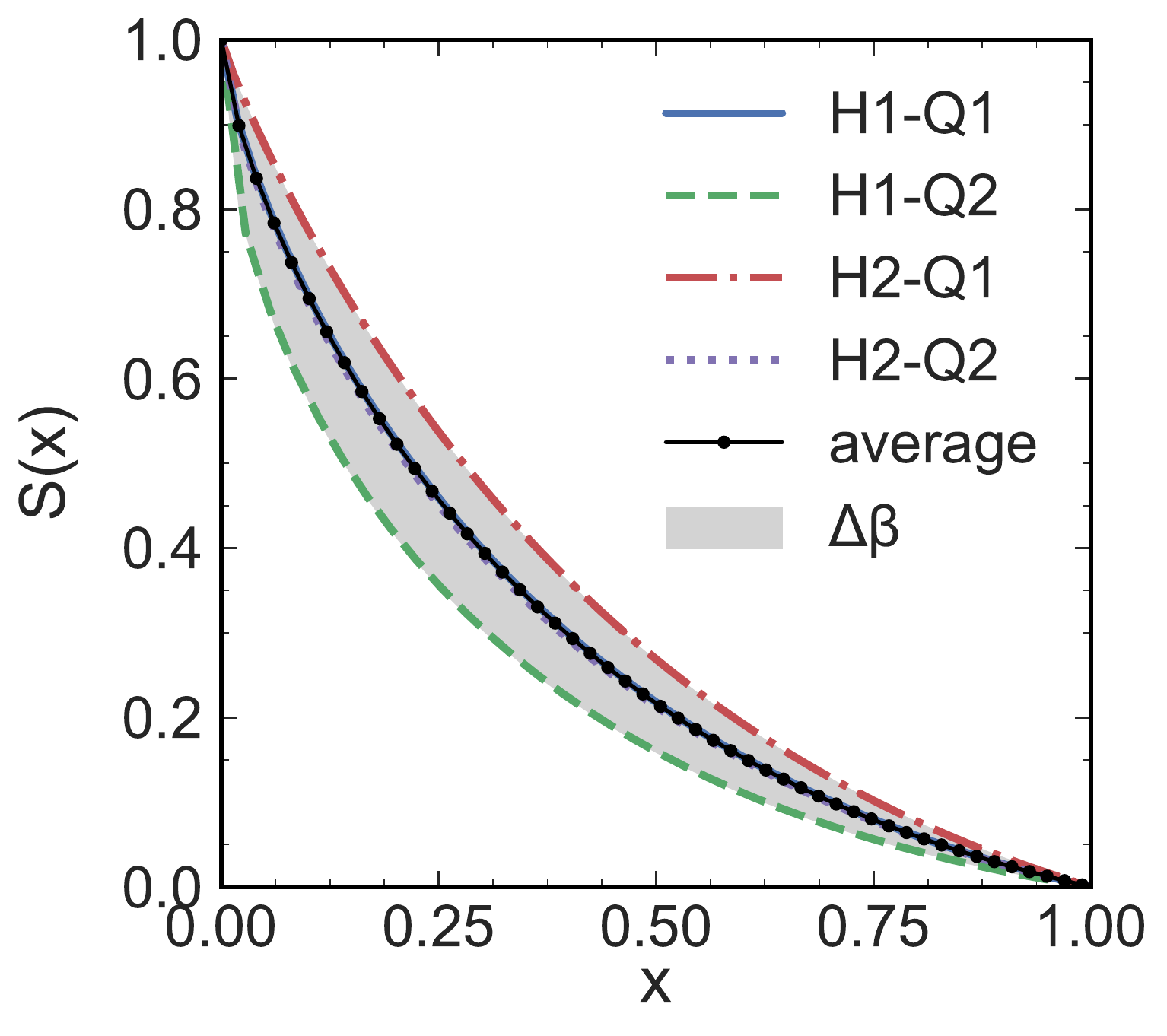}
	\caption{Same as Fig.~\ref{fig::P_sigma} but with inclusion of muons. }
	\label{fig::P_sigma_muons}
\end{figure}

\begin{figure}
	\includegraphics[width=\linewidth]{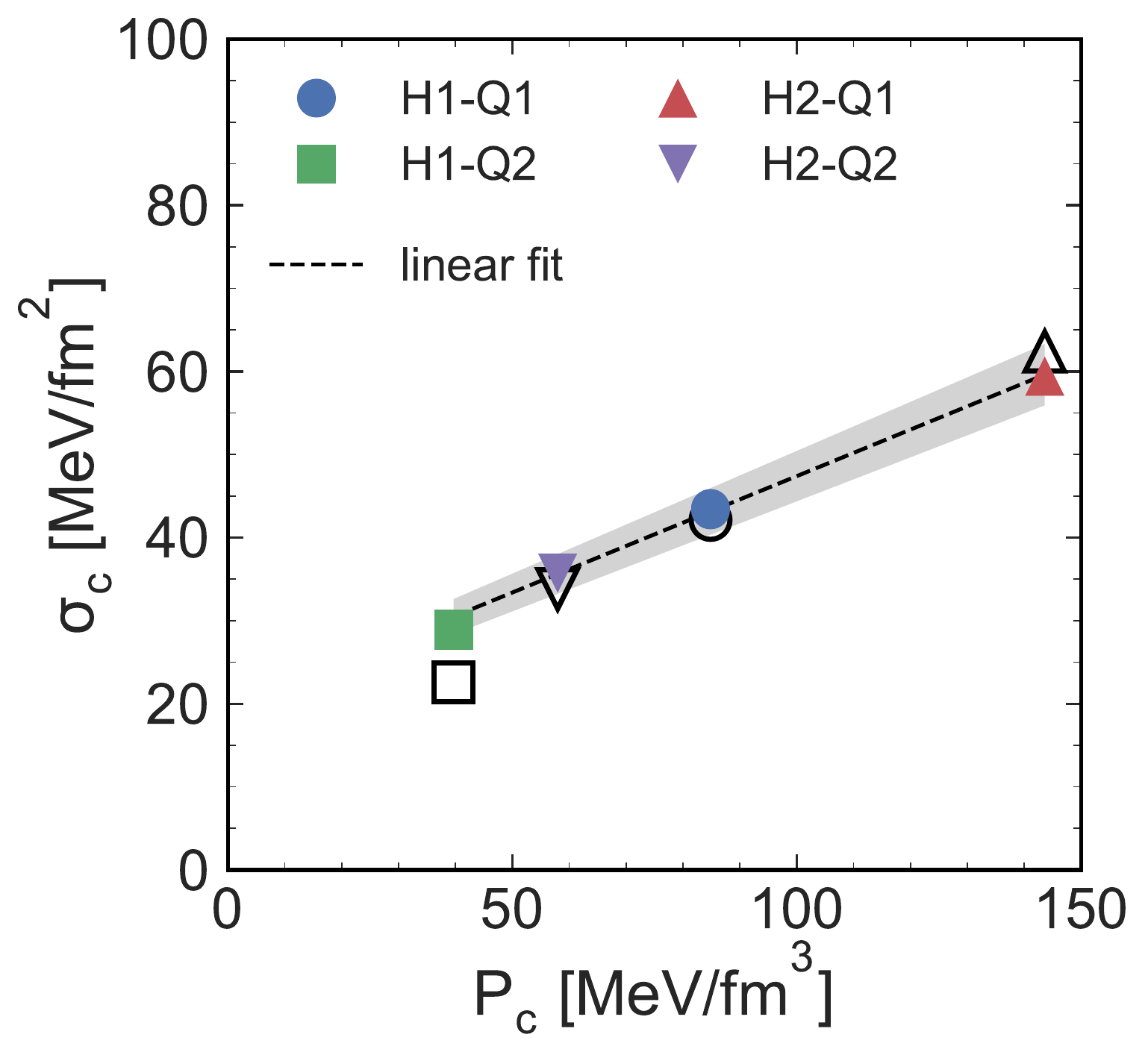}
	\caption{Same as Fig.~\ref{fig::sigma_c_Pc} but with inclusion of muons.}
	\label{fig::sigmac_muons}
\end{figure}

\section{Inclusion of muons}
\label{sec:muons}
The condition of beta equilibrium inevitably leads to an appearance of muons when the electric chemical potential 
reaches the value $\mu_Q = m_\mu = 105\,{\rm MeV}$. 
Their appearance impacts both the EoS and the screening lengths in the medium. 
In \cite{Voskresensky:2001jq,Voskresensky:2002hu,Yasutake:2014oxa} for simplicity only electrons were considered. 
In this section we describe the effect of muons on the properties of the pasta phases.
The corresponding parameters are given in the lower part of Table \ref{tab:Psigma_fit}.

It is known that the influence of muons on the EoS and, in turn, on the properties of compact
stars is rather weak. But, despite that, we see that their impact on the Debye screening length in the hadronic phase is more
pronounced, which affects the values of the critical surface tension.
The dependence of the mixed phase parameter $\Delta_P$ on the surface tension $\sigma$ according to Eq.~(\ref{eq::DeltaP})
is given in the upper panel of Fig.~\ref{fig::P_sigma_muons}, while its scaled form in terms of dimensionless variables according to 
Eq.~(\ref{eq::S(x)})
is shown in the lower panel of that figure. Now the solid line on the lower panel denotes the function 
$S(x; \overline{\beta}')$, with $$\overline{\beta}' = 0.64$$ denoting the mean value of the $\beta$ parameter with the inclusion of muons. As in Fig.~\ref{fig::P_sigma}, the shaded area reflects a symmetric variation of $\beta$ covering all the lines, with an uncertainty now being taken as $\Delta \beta = 0.2$.

The crucial dependence of the critical surface tension $\sigma_c$ in Eq.~(\ref{eq::DeltaP}) on the critical pressure $P_c$ of the Maxwell construction  obeys again Eq.~(\ref{eq::linfit}) 
with the same offset in pressure $P_0$ and the offset in the surface tension 
$\sigma_0=31.6 \pm 1.19$ MeV/fm$^2$
being in accordance within error bars with 
the one found for the case without muons. 
Just the slope $d=0.45 \pm 0.02$ fm is about 5/3 as large as in the case without muons; see Fig.~\ref{fig::sigmac_muons}. This noticeable change in the slope shows that the inclusion of muons is important for further phenomenological studies
of the pasta phases.	

\section{Properties of compact stars}
\label{sec:CS}
Given the EoS for cold, degenerate hybrid star matter one can analyze the properties of the corresponding compact star configurations following from the general relativistic equations of hydrostatic equilibrium. 
Here we follow the analysis step as described, e.g., in Ref.~\cite{Alvarez-Castillo:2018pve} and the literature referred to therein.
In particular, we find the characteristic M-R relationship from solving the Tolman-Oppenheimer-Volkoff (TOV) equations. 
Having obtained the sequence of compact star configurations, one can analyze the moment of inertia $I$ as a function of the 
gravitational mass $M$ of the star. 
In a perturbative treatment, one can analyze the response of the star to tidal deformations, defining the tidal deformability $\Lambda$.  

\begin{figure}
	\includegraphics[width=0.95\linewidth]{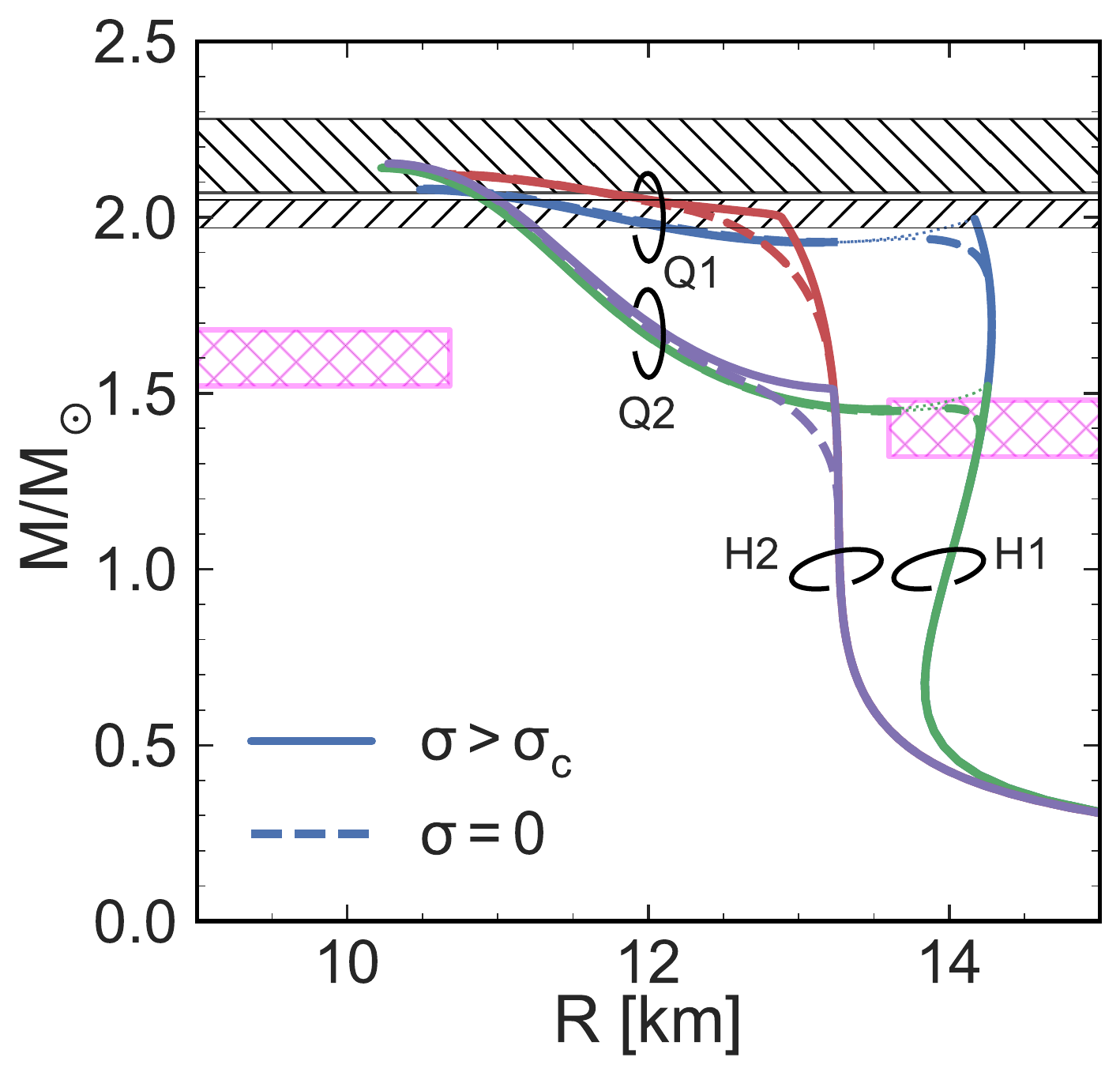}
\vspace{-5mm}
	\caption{M-R relationship for the four pairs of EoSs discussed in this work. Solid lines denote the results for the Maxwell construction ($\sigma>\sigma_c$), and the dashed lines denote the results for $\sigma=0$. Unstable CS configurations are shown by thin dotted lines. 
	The horizontal bands denote the CS mass measurements for PSR J0348+432 
	\cite{Antoniadis:2013pzd} and for PSR J0740+6620 
	\cite{Cromartie:2019kug}.
The magenta cross-hatched regions are excluded by the GW observations from GW170817 with arguments for the left region provided in \cite{Bauswein:2017vtn} and for the right region in \cite{Annala:2017llu}, respectively.
}
	\label{fig::MR}
\end{figure}

In Fig.~\ref{fig::MR} we show the M-R relationship for the pairs of models under consideration for two limiting cases of the mixed phase. Solid lines denote the case with the phase transition described by the Maxwell construction, which is valid for 
$\sigma \gsim \sigma_c$. Dashed lines show the results for the mixed phase described by the construction \eqref{eq::P_grigoryan} with a maximum possible $\Delta_P$ within each model, which in physical terms corresponds to a vanishing surface tension parameter.
The thin dotted lines stand for unstable CS configurations and their presence indicates the existence of a third family branch
for the cases involving the stiff hadronic EoS H1.
Since in this case the third branch of compact stars exists for all $\sigma$ values, it is robust against the formation of pasta structures (which occur for $\sigma<\sigma_c$) for the EoSs of the present work.
We note that for other EoS combinations, e.g., the DD2 EoS for the hadronic phase and the higher order NJL model for the quark matter phase, the third family branches are less robust \cite{Ayriyan:2017nby}. 

The horizontal bands show the  mass range $2.01 \pm 0.04 M_\odot$ measured for pulsar PSR J0348+0432 and the recent 
mass measurement $2.17^{+0.11}_{-0.10} \,M_\odot$  for pulsar PSR J0740+6620  \cite{Cromartie:2019kug}. 
For both models employing the stiffer H1 hadronic EoS there appears a third family of CSs, while both the models based on the softer H2 EoS predict no unstable CS configurations until the maximum CS mass is reached. 
Therefore, within the models under consideration the stiffness of the hadronic EoS essentially determines whether the third family exists or not for a given pair of EoSs. 

The maximum predicted NS masses equal to $[2.08, 2.14] \, M_\odot$ for H1-[Q1,Q2] models and $[2.12, 2.15]\,M_\odot$ for H2-[Q1,Q2] models. All the four hybrid EoSs pass the maximum NS mass constraint, including the recent result of~\cite{Cromartie:2019kug} for the mass of PSR J0740+6620.
The change in the maximum NS mass due to the mixed phase formation is tiny even for the lowest possible surface tension, 
$\sigma= 0$.

\begin{figure}[!thb]
	\includegraphics[width=0.95\linewidth]{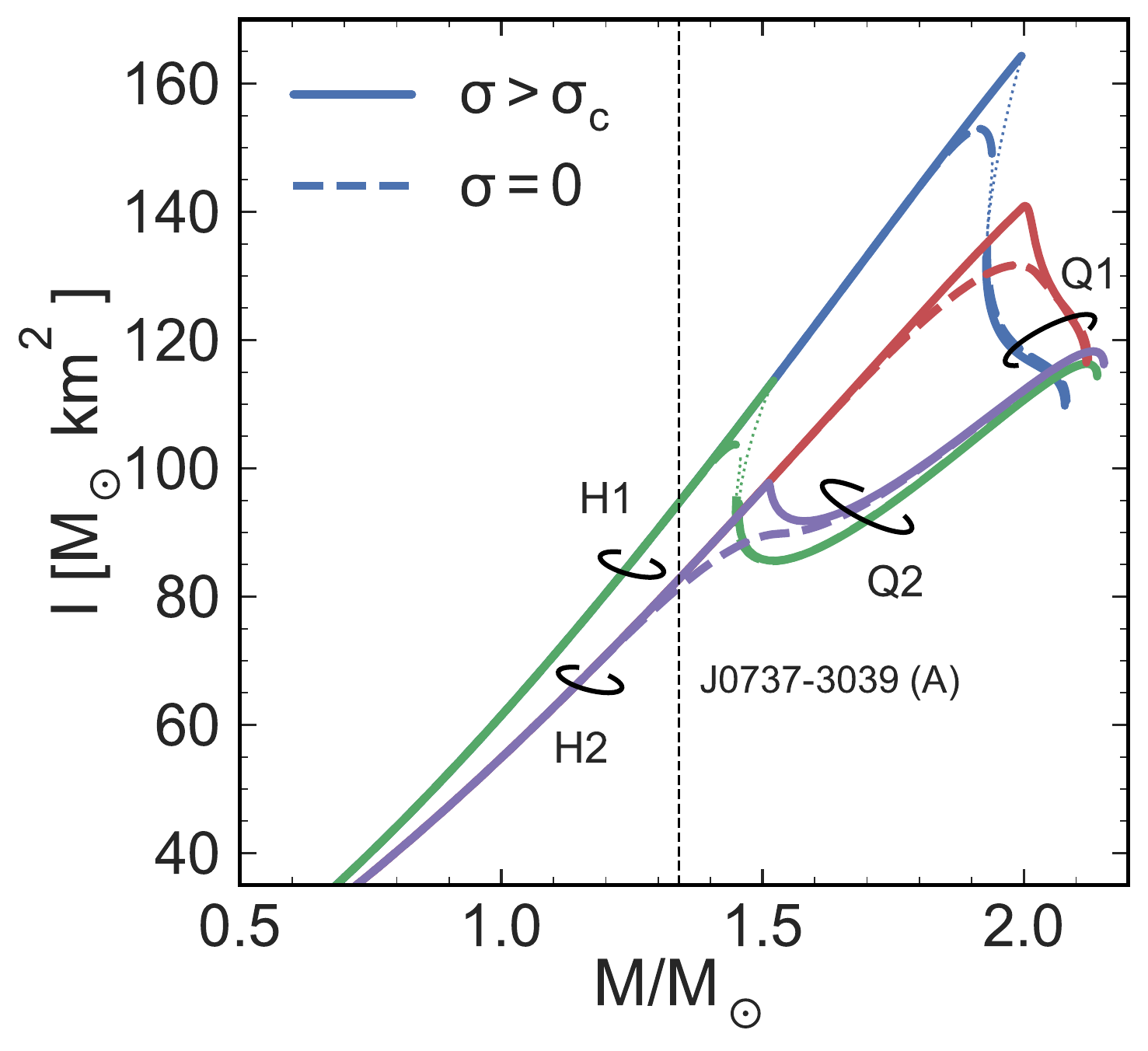}
\vspace{-5mm}
	\caption{Moment of inertia as a function of the compact star gravitational mass. Solid lines show the results for the MC case ($\sigma>\sigma_c$), and dashed lines for $\sigma = 0$. Thin dotted lines denote unstable CS configurations.
	The vertical line indicates the precisely measured mass of star (A) in the binary CS system J0737-3039 
\cite{Kramer:2009zza} for which a determination of the moment of inertia with high precision is under way. }
	\label{fig::IM}
\end{figure}

The predictions for the radius of a canonical CS with $M = 1.4 \, M_\odot$ are 14.2~km for both H1-[Q1,Q2] models and 13.3~km for H2-[Q1,Q2] models in the MC case with $\sigma > \sigma_c$. 
For the H2-Q2 model with a rather low transition mass the radius can be changed by including the mixed phase. 
For the limiting case of $\sigma = 0$ it equals 13.09 km.

        \begin{figure}[!thb]
	\includegraphics[width=0.95\linewidth]{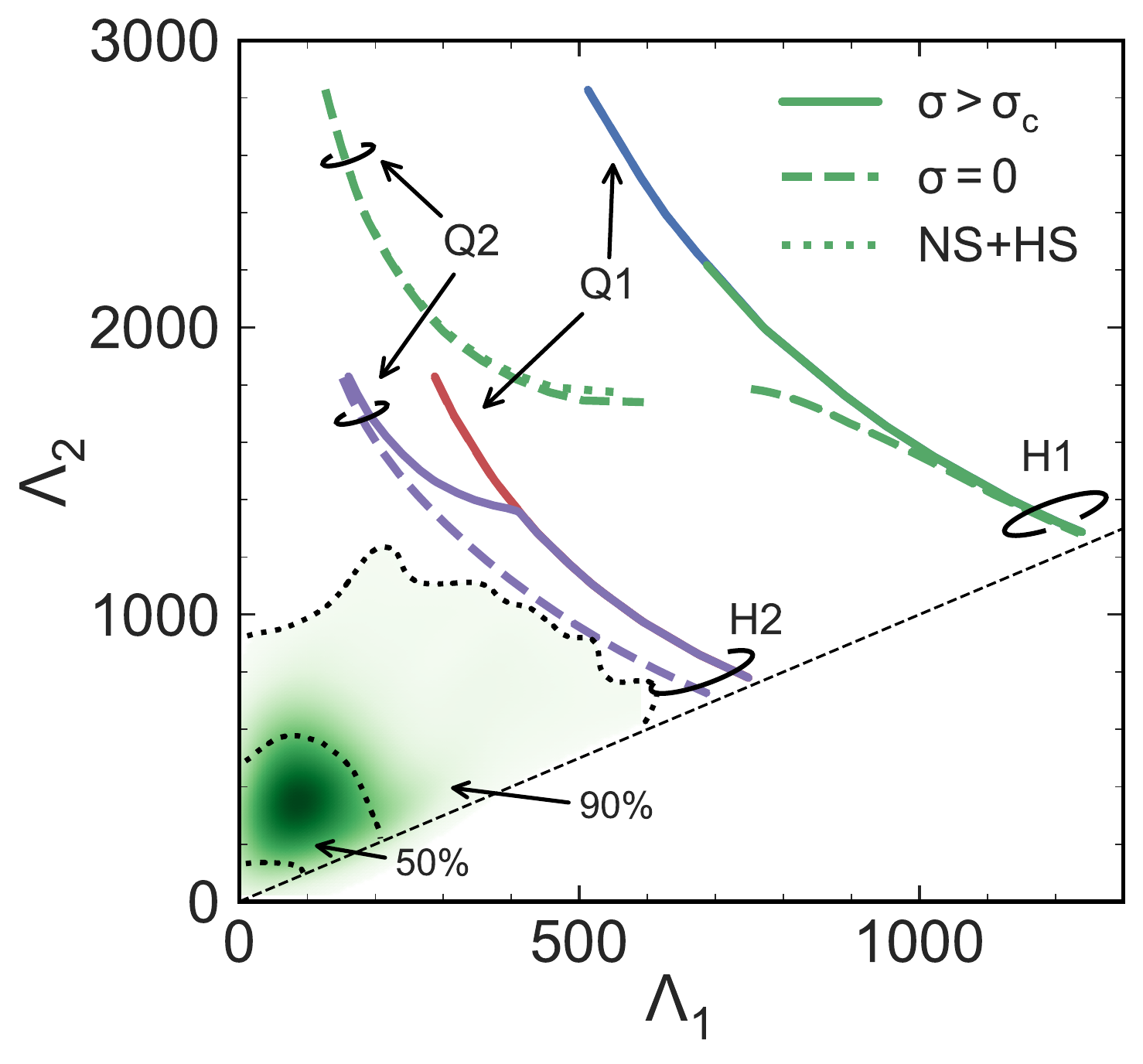}
\vspace{-5mm}
	\caption{ Relation between tidal deformabilities $\Lambda_1$ and $\Lambda_2$ of the two compact stars that merged in the event GW170817 with the chirp mass ${\cal M} = 1.188\,M_\odot$. Solid lines show the results for the MC case ($\sigma>\sigma_c$), and dashed lines -- for $\sigma = 0$. 
	The dotted line shows the NS-HS branch, possible within the H1--Q2 model. 
	For a comparison, the 90\% and 50\% confidence level regions from the analysis of the GW signal by the LIGO-Virgo Collaboration \cite{Abbott:2018exr} are also shown.}
	\label{fig::L1L2}
\end{figure}

Preliminary results for mass and radius of PSR J0030+0451 obtained from an analysis of data taken by the NICER 
experiment within a dual-temperature two-polar-caps model were reported \cite{Guillot:2019} as 
$M=1.44^{+0.17}_{-0.18}\,M_\odot$ and $R=13.84^{+1.18}_{-1.25}$ km.
These values indicate that there might be tension with the mass-radius region excluded by the tidal deformability constraint from GW170817 shown as the right cross-hatched region in that figure.
Such a tension could be resolved by the existence of mass twin stars in that mass interval around $1.4\,M_\odot$, which are a feature accompanying the existence of a third family of compact stars and thus testify for a strong phase transition with a 
large jump of the energy density in compact star matter \cite{Blaschke:2013ana,Benic:2014jia}.
We look forward to the further NICER analyses for this and other millisecond pulsars, with smaller statistical and systematic errors.

In Fig.~\ref{fig::IM} we show the moment of inertia as a function of the compact star mass for the four hybrid EoS cases considered in this work. For orientation, we show the mass of star (A) in the double neutron star system J0737-3039
\cite{Kramer:2009zza}, for which a measurement of the moment of inertia will be accomplished as soon as it reappears as pulsar. Since the measurement of mass and moment of inertia at high accuracy would allow extraction of a precise value of the radius for the same compact object, this would provide a strong constraint for the EoS \cite{Lattimer:2004nj}.

In Fig.~\ref{fig::L1L2} we compare the tidal deformabilities for our set of EoSs with the constraints derived from the observation of GW170817 in the $\Lambda_1-\Lambda_2$ diagram. 
Even with our very restricted set of example EoSs, there are many scenarios possible. The best one among those is that of a binary merger consisting of two hybrid stars (HS) described with the softer hadronic EoS (H2) and an early onset of the deconfinement phase transition via a mixed phase with vanishing surface tension; see  Fig.~\ref{fig::L1L2}. In this case, the hybrid star branch of the CS sequence is connected with the hadronic branch and does not develop a third family.
However, examining  Fig.~\ref{fig::MR} one may speculate that already a small change in the quark matter EoS (Q2), lowering the onset mass of the phase transition by about $0.1\,M_\odot$ would be sufficient to change the situation qualitatively by allowing a HS-HS binary scenario for GW170817 with HS on a rather compact third family branch with much smaller tidal deformability, thus fulfilling the constraint from the LIGO-Virgo Collaboration analysis \cite{Abbott:2018exr} much better.
A systematic investigation of HS EoS in the available parameter space under the available observational constraints will be subject to a Bayesian analysis study that we defer to a subsequent work.

\section{Conclusions}
\label{sec:Conclusions}
	In this work we investigated the modification of quark-hadron hybrid equations of state due to the formation of structures (pasta phases) in the mixed phase. 
	We performed a numerical study of the pasta structures for a set of modern relativistic mean-field equations of state of quark and hadron matter. 
	The surface tension between quark and hadron matter was treated as a free parameter and its value varied in the range 
$\sigma \approx 0-80$ MeV/fm$^2$. 
	We have demonstrated that for all values of $\sigma$ the numerical results for the pressure in the mixed phase 
	can be described reasonably well by a simple polynomial interpolation, which was 
	recently introduced to study the robustness of the occurrence of a third family branch of compact stars against the formation of pasta phases. 
	This finding justifies the application of such a construction to compact stars \cite{Ayriyan:2017nby}. 
	
	The parameter $\Delta_P$ of the construction has the meaning of an additional contribution to the pressure
relative to the critical pressure of the MC, taken at the critical chemical potential of the MC. 
It has its origin in the finite size of the structures in the pasta phases.
As a result of the fit we obtained the functional dependence of $\Delta_P(\sigma)$. This map exhibits the characteristic features of such a phase transition, which we examined quantitatively. 
The curve $\Delta_P(\sigma)$ decreases monotonically with increasing $\sigma$, and its maximum value $\Delta_P(0)$ does not exceed $6 \%$ for any combination of the hadronic and quark matter models considered here. 
In \cite{Ayriyan:2017nby} the upper limit of $\Delta_P$ for existence of a third family of compact stars was found to be of the same order. 
This means that in the realistic case of a non-zero $\sigma$ the third branch of compact stars 
is robust against the formation of pasta structures for practically all values of $\sigma<\sigma_c$ for the EoSs of the present work.
We have confirmed this expectation by explicit calculation of the hybrid star sequences for the examples considered in this work.
Less robust third family branches were obtained for the combination of the DD2 EoS for the hadronic phase and the higher order NJL model for the quark matter phase. 
The actual value of $\Delta_P(0)$ for a given combination of hadronic and quark matter EoSs can be obtained from a fit to the GC of that case.

	It is known that when one accounts for the electric field, there exists a critical value of the surface tension $\sigma_c$, such that for any $\sigma>\sigma_c$ the formation of structures becomes energetically disfavored and the phase transition degenerates to a Maxwell construction with $\Delta_P \simeq 0$. We found the resulting values of  $\sigma_c$ for the models under consideration. 
	Interestingly,  $\sigma_c$ increases monotonically with the critical pressure $P_c$
	of the Maxwell construction for a given pair of hadron and quark  models. 
	We adapted the analytical consideration of \cite{Voskresensky:2002hu} to our models and showed that the analytically evaluated critical surface tension is quantitatively consistent with the numerical results. The existence of the third family requires the phase transition to happen at rather low densities. 
	Therefore, lower values of $P_c$ correspond to models with twin star configurations. Consequently, the critical surface tension for such models should be also lower, and a larger range of $\sigma$ values will not spoil the existence of the third family.
	
With the corresponding fit formula we provide a quantitative relationship that is very useful for practical applications. It allows one to construct the mixed phase equivalent to a full pasta phase calculation for any choice of the quark-hadron surface tension provided a hadronic and a quark matter EoS are given which imply the knowledge of the parameters of both the MC and the GC between them.

We have provided explicit solutions for observable compact star properties for the hybrid EoS examples considered in this work-- M-R relationships, moments of inertia, and tidal deformabilities -- and discuss their present-day constraints from observations.
We have demonstrated the robustness of third family solutions against the formation of pasta phases and illustrated the considerations by a 
conjecture about the possible verification of the existence of third family solutions and thus strong first-order phase transitions with or without pasta structures in the mixed phase.
If NICER will measure a pulsar radius in the mass range of GW170817 which is significantly larger than the upper limit for the radius corresponding to the tidal deformability range deduced from the gravitational wave signal of the inspiral phase of GW170817, then a solution for this puzzle would be the existence of two distinct branches of compact stars in this mass range, leading to "mass twins." 

{\bf Acknowledgments.}
We would like to thank David Alvarez-Castillo and Niels-Uwe Bastian for valuable discussions.
This work has been supported by the Russian Science Foundation under Grant No. 17-12-01427. 
K.M. acknowledges the hospitality of the Chiba Institute of Technology, where a part of this work was done. 
	

\end{document}